\documentclass[letterpaper,twocolumn,10pt]{article}
\usepackage{usenix-2020-09}

\usepackage[pdftex]{graphicx}
\usepackage{longtable}
\usepackage{amsfonts}
\usepackage{amsmath}
\usepackage{amsthm}
\usepackage{enumerate}
\usepackage{pifont}
\usepackage{hyperref}
\usepackage{algpseudocode}
\usepackage{algorithm}
\usepackage{makecell}  
\usepackage{bm}

\graphicspath{{img/}}

\newcounter{margin} 
\DeclareMathOperator*{\argmax}{argmax}
\DeclareMathOperator*{\argmin}{argmin}

\newcommand{\attack}{\texttt{sap}}
\newcommand{\graphm}{\texttt{graphm}}
\newcommand{\liu}{\texttt{freq}}

\newcommand{\nkw}{n}
\newcommand{\ntag}{m}
\newcommand{\nqr}[1]{\eta_{#1}}
\newcommand{\nqravg}{\bar{\eta}}
\newcommand{\Nqr}{\bm{\eta}}

\newcommand{\ndocs}{\ensuremath{N_D}}
\newcommand{\ap}{\mathbf{a}}
\newcommand{\tagg}[1]{\gamma_{#1}}
\newcommand{\kw}[1]{w_{#1}}

\newcommand{\Freqadv}{\tilde{\mathbf{F}}}
\newcommand{\fvecadv}[1]{\tilde{\mathbf{f}}_{#1}}
\newcommand{\freqadv}[2]{\tilde{f}_{#1,#2}}
\newcommand{\Freqobs}{\mathbf{F}}
\newcommand{\fvecobs}[1]{\mathbf{f}_{#1}}
\newcommand{\freqobs}[2]{f_{#1,#2}}
\newcommand{\freqtrue}[2]{{f}^*_{#1,#2}}
\newcommand{\Freqtrue}{\mathbf{F}^*}

\newcommand{\coadv}{\tilde{\mathbf{M}}}
\newcommand{\coobs}{\mathbf{M}}

\newcommand{\voladv}[1]{\tilde{v}_{#1}}
\newcommand{\Voladv}{\tilde{\mathbf{v}}}
\newcommand{\volobs}[1]{v_{#1}}
\newcommand{\Volobs}{\mathbf{v}}
\newcommand{\voltrue}[1]{\bar{v}_{#1}}

\newcommand{\Perm}{\mathbf{P}}
\newcommand{\perm}[1]{p(#1)}

\newcommand{\TPR}{\texttt{TPR}}
\newcommand{\FPR}{\texttt{FPR}}

\begin{document}

\date{}
\author{
{\rm Simon Oya}\\
University of Waterloo
\and
{\rm Florian Kerschbaum}\\
University of Waterloo
} 
\title{\Large \bf Hiding the Access Pattern is Not Enough:\\ Exploiting Search Pattern Leakage in Searchable Encryption}
\maketitle
{\let\thefootnote\relax\footnotetext{\noindent{\textit{To appear in:} Proceedings of the 30th USENIX Security Symposium August 11–13, 2021, Vancouver, B.C., Canada\\
\url{https://www.usenix.org/conference/usenixsecurity21}}}}

\begin{abstract}
Recent Searchable Symmetric Encryption (SSE) schemes enable secure searching over an encrypted database stored in a server while limiting the information leaked to the server.
These schemes focus on hiding the access pattern, which refers to the set of documents that match the client's queries.
This provides protection against current attacks that largely depend on this leakage to succeed.
However, most SSE constructions also leak whether or not two queries aim for the same keyword, also called the \emph{search pattern}.

In this work, we show that search pattern leakage can severely undermine current SSE defenses.
We propose an attack that leverages both access and search pattern leakage, as well as some background and query distribution information, to recover the keywords of the queries performed by the client.
Our attack follows a maximum likelihood estimation approach, and is easy to adapt against SSE defenses that obfuscate the access pattern.
We empirically show that our attack is efficient, it outperforms other proposed attacks, and it completely thwarts two out of the three defenses we evaluate it against, even when these defenses are set to high privacy regimes.
These findings highlight that hiding the search pattern, a feature that most constructions are lacking, is key towards providing practical privacy guarantees in SSE.
\end{abstract}
\section{Introduction}

Searchable Symmetric Encryption (SSE)~\cite{curtmola2011searchable} is a type of private search that allows a client to store an encrypted database in a server while being able to perform searches over it.
In a typical SSE scheme, the client first encrypts the database using private-key encryption, generates a search index, and sends them to the server.
Then, the client can perform queries by generating query tokens, that the server evaluates in the index to obtain which documents match the query.

There are different types of private search techniques that provide different security guarantees and query functionalities, such as range or SQL queries.
Fuller et al.~\cite{fuller2017sok} give an overview of protected search schemes and examples of companies that offer products with searchable encryption.
In this work, we focus on point queries, which are the main query type in SSE schemes.
Namely, we consider that each document in the database has a list of \emph{keywords} associated with it, and the client queries for documents that match a certain keyword.
The typical use case of keyword searches in related work are email databases~\cite{islam2012access,cash2015leakage,zhang2016all,pouliot2016shadow,liu2014search}.

Even though the database and the query tokens are encrypted, basic SSE schemes leak certain information to the server when performing a query.
There are two main sources of leakage considered in the literature: the \emph{access pattern}, which refers to the identifiers of the documents that match a query; and the \emph{search pattern}, also known as query pattern, which refers to identifying which queries in a sequence are identical.
An honest-but-curious server can leverage this leakage to identify the client's queries (query recovery attacks) or the database contents (database recovery attacks).

Liu et al.~\cite{liu2014search} proposed one of the few attacks that exploits only search pattern leakage to recover queries.
The search pattern allows the adversary to compute the \emph{frequency} with which the client performs each query.
After observing queries for a long time, the attacker can compare the frequency information of each query token with auxiliary data to identify each query's keyword.
Islam et al.~\cite{islam2012access} proposed an attack (IKK) that leverages keyword co-occurrence information extracted from the access pattern leakage, as well as certain ground truth information about the client's queries, to identify the remaining queries.
Further refinements of this idea improve the attack when the keyword universe is large~\cite{cash2015leakage} and even allow the adversary to infer the keywords without ground truth and with imperfect auxiliary information~\cite{pouliot2016shadow}.

In order to protect the client against these attacks, the research community has proposed \emph{privacy-preserving SSE schemes} with reduced leakage.
Schemes that completely hide the search pattern, such as those based on Oblivious RAM (ORAM)~\cite{garg2016tworam}, require running a protocol with a typically prohibitive communication cost.
Also, they still leak the \emph{response volume}, i.e., how many documents are returned in response to a query, which can be exploited by certain attacks~\cite{cash2015leakage}.

Recent proposals trade in communication or computational efficiency for privacy.
Some of these defenses propose relaxations of the notion of ORAM~\cite{demertzis2020seal}, or simply obfuscate the access pattern by adding false positives and false negatives to the set of documents that match a query~\cite{chen2018differentially}.
Recent work by Patel et al.~\cite{patel2019mitigating} proposes using hashing techniques to completely obfuscate the access pattern structure, and hide the response volume by padding it with Laplacian noise.

The privacy guarantees of these and other defenses can be assessed theoretically or empirically.
Theoretical notions include the differential privacy framework~\cite{dwork2008differential}, used to protect access pattern leakage~\cite{chen2018differentially} or response volume~\cite{patel2019mitigating}, or quantifying the number of information bits revealed per query~\cite{demertzis2020seal}.
The problem with these theoretical notions is that it is hard to judge how well they translate into actual protection guarantees against attacks.
Assessing the performance of defenses empirically using generic SSE attacks can however overestimate the protection of these defenses.
Most works either evaluate their proposals against ad-hoc attacks~\cite{demertzis2020seal}, figure out how to extend existing attacks to a given defense (e.g., Chen et al.~\cite{chen2018differentially} adapt IKK~\cite{islam2012access}), or simply rely only on a theoretical guarantee~\cite{patel2019mitigating}.
The effectiveness of current defenses has only been evaluated against attacks that exploit access pattern leakage, while search pattern leakage has only recently been explored in the particular case of range and nearest-neighbor queries~\cite{kornaropoulos2020state}.

In this work, we aim at investigating to which extent leaking the search pattern affects the privacy of SSE schemes that allow point queries.
In order to achieve this, we propose the first query identification attack that simultaneously combines access and search pattern leakage, as well as some auxiliary (background) information, to identify the keywords of the client's queries.
We note that, even though certain attacks rely on strong background information~\cite{islam2012access, cash2015leakage} to achieve high accuracy~\cite{blackstone2020revisiting}, our assumptions on background information are weak.
For example, we do not assume that the adversary knows the true distribution of the documents/keywords nor any ground-truth information.
Instead of relying on heuristics, we develop our attack following a Maximum Likelihood Estimation (MLE) approach.
This makes our attack easy to adapt against specific defenses, and we illustrate this by modifying our attack to perform well against three of the most recent privacy-preserving SSE schemes for point queries \cite{chen2018differentially,patel2019mitigating,demertzis2020seal}.

We compare our attack with the state-of-the-art graph matching attack by Pouliot and Wright~\cite{pouliot2016shadow}, and show that our proposal is orders of magnitude faster and has a higher query recovery accuracy than graph matching when the client does not query for every possible keyword in the dataset.
Our attack also outperforms one of the few attack that uses search pattern leakage~\cite{liu2014search}.
The main reason that our attack outperforms previous works is that it \emph{combines} volume and frequency leakage information.
Our attack achieves $74\%$, $48\%$, $37\%$, and $22\%$ query recovery rate for keyword universes of sizes $100$, $500$, $1\,000$, and $3\,000$, respectively, after observing only $\approx 250$ (possibly repeated) queries from the client.

We tune our attack against three recent privacy-preserving SSE schemes~\cite{chen2018differentially,patel2019mitigating,demertzis2020seal} and evaluate its performance with two real datasets.
Our experiments reveal that these defenses are highly effective against a naive attack that does not take the defense into account (e.g., lowering the accuracy with $1\,000$ possible keywords from $37\%$ to $1.4\%$, $2.4\%$, and $2.7\%$ respectively for defenses \cite{chen2018differentially}, \cite{patel2019mitigating}, and \cite{demertzis2020seal}, configured to high privacy regimes).
When adapting our attack against the defenses, the accuracy increases back to $30\%$, $35\%$, and $23\%$, respectively.
This shows that two of the defenses fail at achieving meaningful protection levels even though they incur more than $400\%$ communication overhead.
The third defense~\cite{demertzis2020seal} is both more efficient and effective, but our attack still recovers a non-trivial amount of keywords against it.

To summarize, our contributions are:
\begin{enumerate}
	\item We derive a new query recovery attack for SSE schemes following a maximum likelihood estimation approach.
	Our attack combines information from both access and search pattern leakage.
	\item We evaluate our attack against a basic SSE scheme and show that it is more accurate than the state-of-the-art access pattern-based attack and one of the few attacks that relies exclusively on search pattern leakage.
	\item We provide a methodology to adapt our attack against particular SSE defenses and illustrate our approach by tailoring our attack to perform well against three recent proposals.
	\item We evaluate our attack against these three defenses and show that two of them in practice fail to protect the queries and we still recover a non-trivial amount of queries against the third one.
\end{enumerate}

The rest of the paper is organized as follows. 
We summarize related work in the next section.
In Section~\ref{sec:pre} we introduce our general leakage model for SSE schemes that we use to derive our attack in Section~\ref{sec:att} and adapt it against defenses in Section~\ref{sec:defenses}.
We compare our attack with others and evaluate it against SSE defenses in Section~\ref{sec:eval}, discuss how to hide search pattern leakage in Section~\ref{sec:discussion} and conclude in Section~\ref{sec:conclusions}.

\section{Related Work}
\label{sec:realwork}

Searchable Symmetric Encryption (SSE)~\cite{curtmola2011searchable} is one type of \emph{protected search} technique.
Other popular protected search techniques include Property-Preserving Encrpytion (PPE)~\cite{pandey2012property} and Privacy Information Retrieval (PIR)~\cite{chor1995private}.
We refer to the SoK paper by Fuller et al.~\cite{fuller2017sok} for a thorough revision of these and other protected database search techniques.
In this section, we summarize the main attacks and defenses in SSE, with a particular focus on point queries, which is the subject of our work.

\subsection{Attacks against SSE Schemes}

Attacks against SSE schemes can be broadly classified based on whether they consider an active or passive adversary, the type of queries allowed by the scheme, the leakage required by the attack, and the goal of the attack.

File injection attacks~\cite{cash2015leakage,zhang2016all} consider an \emph{active adversary} that is able to insert documents in the database.
This is reasonable, for example, if the database is an email dataset and the adversary can send emails to be stored in that dataset.
By carefully choosing the keywords of the inserted documents and studying which of these files match a certain query, the adversary can identify the underlying keyword of such query.

We can broadly classify \emph{passive attacks} according to their goal into database and query recovery attacks.
Database recovery attacks aim to recover the content of the database, while query recovery attacks aim to find the target of each of the client's queries.
In some schemes, query recovery attacks can be used to recover the contents of the database by checking which queries trigger a match for each document.

Database recovery is a typical goal of attacks in \emph{range query} schemes.
In these schemes, each document has a particular attribute value and the client can retrieve documents whose attribute is within a given range.
Previous works study the complexity of recovering the attribute values in the dataset based on the access pattern leakage of range queries~\cite{kellaris2016generic, lacharite2018improved, grubbs2019learning,gui2019encrypted}.
Recent work by Kornaropoulos et al.~\cite{kornaropoulos2020state} also uses the search pattern leakage (i.e., whether or not two queries are identical) to develop reconstruction attacks for range and $k$-nearest neighbor query schemes. 
These works are not necessarily relevant for our work, since they require schemes that allow range queries.

Query recovery is a typical goal of attacks against SSE schemes where the client performs \emph{point queries}, i.e., it queries for the set of documents that contain a certain keyword.
In this setting, we can generally distinguish between attacks that use access pattern leakage and those that use search pattern leakage.

The seminal work by Islam et al.~\cite{islam2012access} (known as IKK attack) shows that it is possible to recover the client's queries using \emph{access pattern} leakage, but relies on strong assumptions on background information.
In this attack, the adversary stores how many documents match every pair of distinct queries and compares this with auxiliary information about keyword co-occurrence.
Then, it matches each received query with a keyword using a heuristic algorithm that also relies on ground truth information about a subset of the queries.
Cash et al.~\cite{cash2015leakage} showed that IKK does not perform well when the subset of possible keywords is large (e.g., $2\,500$ keywords) and propose an alternative attack that identifies keywords based on their response volume (i.e., the number of documents that match the query).
The most recent iteration of these attacks, by Pouliot and Wright~\cite{pouliot2016shadow}, proposes a graph matching attack that allows the adversary to accurately recover the queries even when the adversary has imperfect auxiliary information about the statistical distribution of the dataset.

The attack proposed by Liu et al.~\cite{liu2014search} relies only \emph{search pattern} leakage.
This attack assigns a tag to each distinct query it receives, and uses the search pattern leakage to monitor the frequency of each tag over time.
Then, the adversary can recover the underlying keyword of each tag by comparing the tag query trends with keyword trend information.

Ours is the first attack against SSE schemes where the client performs point queries that leverages \emph{both access and search pattern} leakage.
Our attack takes core ideas from related works \cite{pouliot2016shadow,liu2014search}, but relies on a Maximum Likelihood Estimation (MLE) approach to find the most likely keyword of each received query.
The techniques we use to solve our attack are somewhat similar to the frequency-based database recovery attacks by Bindschaedler et al.~\cite{bindschaedler2018tao} in deterministic encryption.
However, our adversary model is conceptually very different since it aims at query recovery, and our attack leverages both frequency and volume (search pattern) information.

\subsection{Privacy-Preserving SSE Schemes}

Early works that introduce attacks against SSE schemes also propose the first techniques to partially hide access pattern information~\cite{islam2012access} or query frequencies~\cite{liu2014search} to palliate the effects of these attacks.
Even though one can build protected search techniques based on Oblivious RAM (ORAM)~\cite{goldreich1996software} that completely hide the search pattern (and possibly the access pattern), such as TwoRAM~\cite{garg2016tworam}, their practicality is still questionable since they incur a significant communication overhead and they still leak the query volume information.
Kamara et al.~\cite{kamara2018structured} provide a framework to design structured encryption schemes while hiding the access and search pattern. 
Their approach is based on the square-root ORAM by Goldreich and Ostrovsky~\cite{goldreich1996software}, and introduces the notion of volume-hiding encrypted multimap schemes to hide the volume information (e.g., how many documents are associated with every search key).
Patel et al.~\cite{patel2019mitigating} propose more efficient volume-hiding techniques.
They explain why completely hiding the query response volume is unreasonably expensive, and introduce differentially-private volume-hiding, which trades leakage for efficiency.

Chen et al.~\cite{chen2018differentially} propose a framework to hide access patterns in a differentially private way.
In their scheme, the client first generates an inverted index, i.e., a structure indicating which documents contain which keywords, and obfuscates it by adding false positives and false negatives.
This obfuscation adds noise to the access patterns and thus makes it harder to apply attacks such as IKK~\cite{islam2012access} against it.
They palliate false positives by using a document redundancy technique.

Finally, recent work by Demertzis et al.~\cite{demertzis2020seal} proposes an ORAM-based scheme with the idea of hiding bits of information about the address of a document in the database and the response volume of a query.
For this, they split the dataset into $2^\alpha$ ORAM blocks that hide which document within the block is accessed each time, and pad the response volume of each query to the next power of a constant $x$.
The values of $\alpha$ and $x$ allow to adjust the privacy vs.~utility trade-off of this scheme.

\section{Preliminaries}
\label{sec:pre}

We consider a client-server scenario where the client owns a database and, for the sake of saving storage space, wants to outsource it to the server while keeping the ability to perform \emph{point} queries over it. 
The client uses a (privacy-preserving) SSE scheme for this, that works as follows.
First, the client encrypts the database using symmetric encryption and sends it to the server, together with a query index. 
Then, when the client wants to query for a particular keyword, it generates a query token and sends it to the server. 
The server evaluates the query token on the index and obtains the addresses of the documents that match the query. 
The server returns these documents to the client.
The client wants to keep both the underlying keyword of each query and the contents of the database secret (keyword and database privacy).

The adversary that we consider is an honest-but-curious server that follows the protocol but might use the information it observes to infer private information.
Throughout the text, we refer to the server as adversary or attacker.
We focus on query recovery attacks, i.e., the goal of the adversary is to identify the underlying keyword behind each query.
In some cases, the adversary can leverage query recovery attacks to recover the database by identifying the set of keywords that trigger a match for each document in the database.
We always assume that the adversary knows the parameters and algorithms of the SSE scheme, following Kerckhoffs' principle.

\begin{table}[t]
\centering
\begin{tabular}{ r  l }
\hline
\multicolumn{2}{c}{General Parameters} \\ \hline
$\Delta$ & Keyword universe $\Delta\doteq[\kw{1},\kw{2},\dots,\kw{\nkw}]$. \\
$\nkw$ & Total number of keywords, $\nkw\doteq|\Delta|$.\\
$\kw{i}$ & $i$th keyword, with $i\in[\nkw]$.\\
$\ndocs$ & Number of documents in the encrypted dataset.\\
$\rho$ & Number of observation time intervals. \\ \hline

\multicolumn{2}{c}{Adversary Observations} \\ \hline
$\ntag$ & Number of tags (distinct access patterns observed).\\
$\tagg{j}$ & $j$th tag, with $j\in[\ntag]$.\\
$\ap_j$ & Access pattern assigned to tag $j$.\\
$\volobs{j}$ & Volume of a query with tag $j$, $\volobs{j}\doteq|\ap_j|$.\\
$\Volobs$ & Volume of tags, $\Volobs\doteq[\volobs{1},\dots,\volobs{\ntag}]$.\\
$\coobs$ & Tag co-occurrence matrix (size $\ntag\times\ntag$).\\
$\nqr{k}$ & Number of queries sent in the $k$th time interval.\\
$\Nqr$ & Vector $\Nqr\doteq[\nqr{1}, \nqr{2}, \dots, \nqr{\rho}]$.\\
$\freqobs{j}{k}$ & Query frequency of $\tagg{j}$ in the $k$th time interval.\\
$\fvecobs{j}$ & Query frequency vector of $\tagg{j}$, $\fvecobs{j}\doteq[\freqobs{j}{1}, \dots, \freqobs{j}{\rho}]$.\\
$\Freqobs$ & Query frequency matrix of all tags (size $\ntag\times\rho$).\\ \hline

\multicolumn{2}{c}{Auxiliary (Background) Information} \\ \hline
$\voladv{i}$ & Auxiliary volume information for keyword $\kw{i}$.\\
$\Voladv$ & Volume vector of keywords, $\Voladv\doteq[\voladv{1},\dots,\voladv{\nkw}]$.\\
$\coadv$ & Auxiliary keyword co-occurrence matrix ($\nkw\times\nkw$).\\
$\freqadv{i}{k}$ & Query frequency of $\kw{i}$ in the $k$th time interval.\\
$\fvecadv{i}$ & Query frequency vector of $\kw{i}$, $\fvecadv{i}\doteq[\freqadv{i}{1}, \dots, \freqadv{i}{\rho}]$.\\
$\Freqadv$ & Query frequency matrix of all keywords (size $\nkw\times\rho$).\\ \hline

\multicolumn{2}{c}{Attack Goal} \\ \hline
$\perm{j}$ & Index of the keyword that the attack assigns to $\tagg{j}$.\\
$\Perm$ & Permutation matrix, $\Perm_{\perm{j},j}=1$, else 0 ($\nkw\times\ntag$).\\ 
\end{tabular}
\caption{Summary of notation \label{tab:notation}}
\end{table}

\begin{figure}
\begin{center}
\def\svgwidth{\linewidth} 
{
\fontsize{7.5pt}{9.5pt}
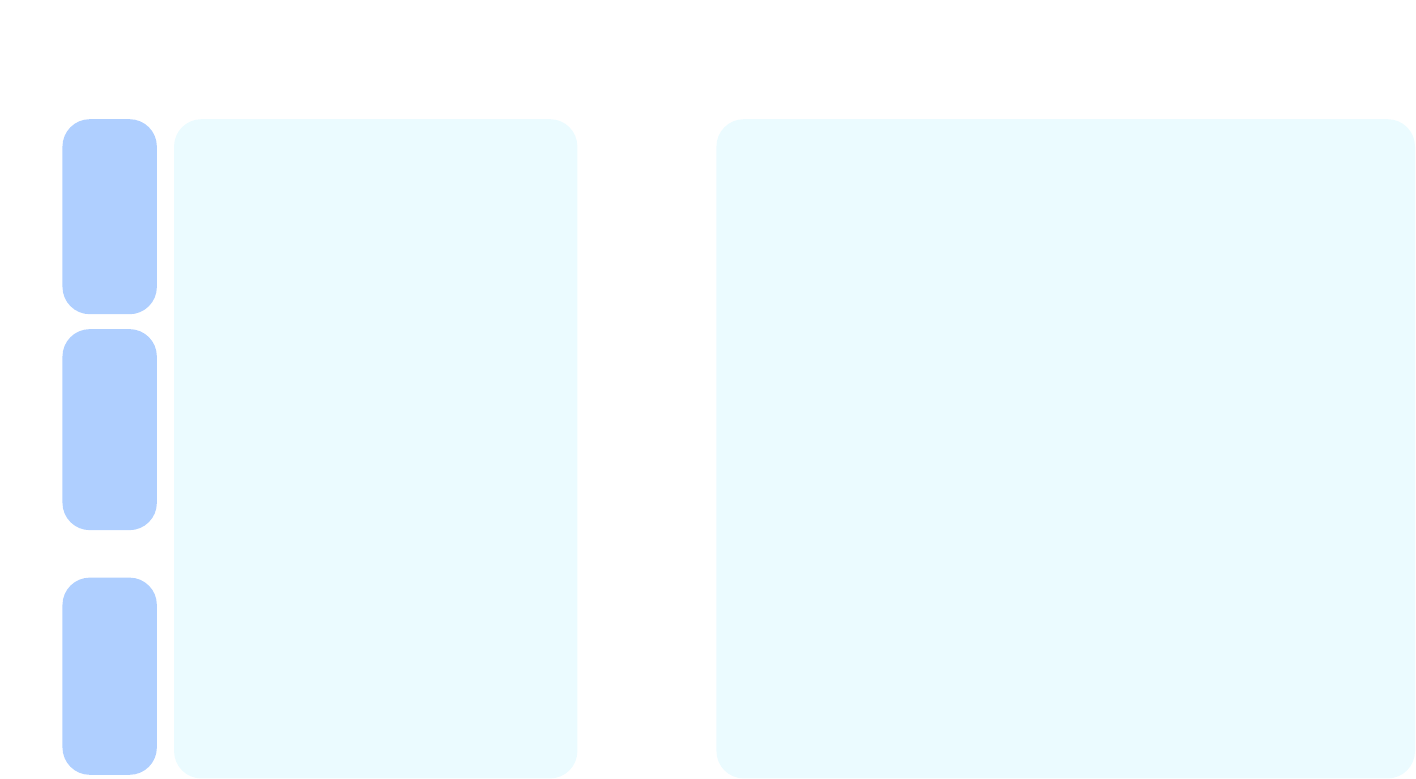
}
\caption{System Model \label{fig:model}}
\end{center}
\end{figure}

\subsection{System Model and Notation}
We present a general model that captures the leakage of many proposed privacy-preserving SSE schemes while abstracting from the cryptographic and implementation details of these protocols.
The notation that we use is summarized in Table~\ref{tab:notation}.
We use upper-case boldface characters to denote matrices and lower-case boldface characters to denote vectors.
The $(i,j)$th entry of matrix $\mathbf{A}$ is $(\mathbf{A})_{i,j}$, and $\text{tr}(\mathbf{A})$ is the trace of $\mathbf{A}$.
We represent the natural logarithm as $\log$; other logarithm bases are written explicitly.

Let $\Delta=[\kw{1}, \kw{2}, \dots, \kw{\nkw}]$ be the keyword universe, where $\kw{i}$ is the $i$th keyword, and let $\nkw\doteq|\Delta|$ be the total number of keywords. 
Let $\ndocs$ be the number of documents in the encrypted database that the client sends to the server.
For each query, the adversary observes the tuple $(t, \ap)$ where $t$ is the timestamp of the query and $\ap$ is the \emph{access pattern}, i.e., a vector with the positions of the documents that match the query.
The leakage of all the SSE schemes that we consider in this work can be characterized by a sequence of tuples $(t,\ap)$.
We use $|\ap|$ to denote the \emph{response volume}, i.e., the number of documents returned to the client in response to a query.

We consider SSE schemes that leak the \emph{search pattern}, i.e., they leak which queries within a sequence are for the same keyword.
The search pattern leakage can be explicit or implicit.
Explicit search pattern occurs when querying for a certain keyword always generates the same query token~\cite{curtmola2011searchable, chen2018differentially, patel2019mitigating}.
Implicit leakage refers to SSE schemes where the queries for the same keyword $\kw{i}$ always generate the same access pattern $\ap$, and the adversary can compare access patterns to check whether or not different tokens aim for the same keyword~\cite{demertzis2020seal}.
We discuss how to hide search patterns in Section~\ref{sec:discussion}.

Using the search pattern leakage, the adversary can assign a \emph{tag} to each different access pattern it observes.
The number of tags $\ntag$ will be at most equal to the number of keywords $\nkw$ (i.e., $\ntag\leq\nkw$), and will be strictly smaller if the client does not query for all possible keywords during the observation time.
We use $\ap_j$ to denote the access pattern of the $j$th tag, with $j\in[\ntag]$.
Then, the goal of the query recovery attack is to assign each tag its correct keyword.
We denote this assignment, which is an injective mapping, by $\perm{\cdot}: [\ntag]\to[\nkw]$.
We also represent it in matricial form as a $(\nkw\times\ntag)$ permutation (column-selection) matrix that we denote by $\Perm$ and define as
\begin{equation}
	(\Perm)_{i,j}=\begin{cases} 1\,, & \text{if }i=\perm{j}\,,\\
	0\,, & \text{otherwise.}
	\end{cases}
\end{equation}

Figure~\ref{fig:model} illustrates this model and notation.
In the figure, the client queries for keywords $\kw{12}, \kw{23}, \kw{51}, \dots,\kw{12}$.
The server evaluates the query tokens in the search index and obtains which documents in the encrypted database match each query (i.e., the observed access patterns).
Then, the server assigns a tag $\tagg{j}$ to each distinct access pattern.
Note that the access patterns that result from evaluating different query tokens generated from the same keyword (e.g., $\kw{12}$) are identical.
The goal of the attack is to map each $\tagg{j}$ to a keyword $\kw{i}$.
In order to perform this mapping, the server uses information from the structure of the access patterns and from the frequency with which the server observes each access pattern, as well as some auxiliary information that we specify below.

Below, we define different data structures that the adversary can compute from the observations.
Several query recovery attacks~\cite{islam2012access, pouliot2016shadow, liu2014search}, as well as our proposal, can be defined by using these variables.
The following structures are computed from the access patterns:
\begin{itemize}
	\item \textbf{Query volume} ($\Volobs$, $\volobs{j}$).
	The query volume refers to the number of documents in the database that are returned as a response to a certain query.
	We use $\volobs{j}\in[0,1]$ to denote the normalized volume of the $j$th tag, i.e., $\volobs{j}\doteq|\ap_j|/\ndocs$, and $\Volobs\doteq[\volobs{1}, \dots, \volobs{\ntag}]$.
		
	\item \textbf{Co-occurence matrix} ($\coobs$).
	This variable refers to the number of documents that simultaneously match two different queries, normalized by the total number of documents in the database.
	We use $\coobs$ to denote the symmetric matrix whose $(i,j)$th element is $(\coobs)_{i,j}\doteq|\ap_i \cap \ap_j|/\ndocs\in[0,1]$.
\end{itemize}	

The following structures are computed from the search patterns, i.e., from how many times the client sends a query tagged as $\tagg{j}$.
In order to compute these structures, the adversary first splits the observation time into $\rho$ intervals (e.g., weeks).

\begin{itemize}
	\item \textbf{Query number} ($\Nqr$, $\nqr{k}$).
	We use $\nqr{k}$ to denote the number of queries the client sent in the $k$th interval, and define the vector $\Nqr\doteq[\nqr{1},\dots,\nqr{\rho}]$.
	\item \textbf{Query frequency} ($\Freqobs, \fvecobs{j}, \freqobs{j}{k}$).
	The query frequency refers to how often the client performs a certain query.
	For each tag $\tagg{j}$ ($j\in[\ntag]$) and each time interval, indexed by $k\in[\rho]$, we use $\freqobs{j}{k}$ to denote the frequency of tag $j$ in the $k$th interval, i.e., the total number of times the client queries for tag $j$ in the interval, divided by the total number of queries in that interval.
We use $\fvecobs{j}$ to denote the vector that stores $\freqobs{j}{k}$ for all $k\in[\rho]$ and $\Freqobs$ is the $(\ntag\times \rho)$ matrix that stores all the frequencies.
\end{itemize}

In addition to the observations, the adversary has certain \emph{auxiliary background information} (e.g., a training set) that helps them carrying out the query recovery attack.
The adversary uses this information to compute data structures like the ones defined above, but for each keyword instead of each tag.
We denote the auxiliary query volume information by $\voladv{i}$ for each keyword $i\in[\nkw]$, the $n\times n$ co-occurrence matrix of keywords by $\coadv$, and the $\nkw\times\rho$ matrix storing the query trends of each keyword by $\Freqadv$.
We note that background information is a strong assumption and attacks that rely on high-quality auxiliary information to be effective might be unrealistic~\cite{blackstone2020revisiting}.
In our evaluation in Section~\ref{sec:eval}, we show that our attack is strong under weak assumptions on the auxiliary information.
Namely, in our experiments the adversary computes $\Voladv$ and $\coadv$ using a training set that is disjoint with the actual client's database, and $\Freqadv$ using public information about query trends with a time offset.

Below, we explain state-of-the-art query recovery attacks using access pattern~\cite{pouliot2016shadow} and search pattern~\cite{liu2014search} leakage using our notation.

\subsection{Graph Matching Attack}
\label{sec:graphm}

In the graph matching attack by Pouliot and Wright~\cite{pouliot2016shadow}, the adversary represents the set of tags and the set of keywords as two graphs, and the goal is to solve a \emph{labeled graph matching} problem between the graphs.
Let the keyword graph be $\tilde{G}$ (it has $\nkw$ nodes), and let the tag graph be $G$ (it has $\ntag$ nodes).
The labeled graph matching problem looks for the permutation matrix $\Perm$ that minimizes the convex combination of two objective functions that measure a similarity score between the graphs.

The first objective function is based on the adjacency matrices of each graph, that determine the weights of the edges between nodes. The adjacency matrix of $\tilde{G}$ is $\coadv$, and the adjacency matrix of $G$ is $\coobs$.
Given an assignment of keywords to tags $\Perm$, the adjacency matrix of an upscaling of $G$ to match the size of $\tilde{G}$ would be $\Perm\coobs\Perm^T$.
Therefore, it makes sense to look for the permutation $\Perm$ that minimizes
\begin{equation}
	||\coadv-\Perm\coobs\Perm^T||_F^2\,,
\end{equation}
where $||\cdot||_F$ denotes the Frobenius norm of matrices.\footnote{The original attack~\cite{pouliot2016shadow} considers the Frobenius (or Euclidean) norm, but the software package that they use to solve the problem~\cite{zaslavskiy2008path} uses the Frobenius norm \emph{squared}.}

Additionally, the labeled graph matching attack considers another objective function that depends only on the volume of each keyword/tag.
The attack builds a $\nkw\times\ntag$ similarity matrix $\mathbf{C}$ whose $(i,j)$th element measures the likelihood of the assignment of $\tagg{j}$ to keyword $\kw{i}$.
Pouliot and Wright~\cite{pouliot2016shadow} compute this likelihood assuming that the number of matches of a certain keyword $\kw{i}$ in the encrypted dataset follows a Binomial distribution with $\ndocs$ trials (dataset size) and a match probability given by the volume of that keyword in the auxiliary information $\voladv{i}$.
Then, the $(i,j)$th element of $\mathbf{C}$ is
\begin{equation} \label{eq:volbino}
	(\mathbf{C})_{i,j}={\ndocs \choose \ndocs \volobs{j}}\cdot \voladv{i}^{\ndocs \volobs{j}} (1-\voladv{i})^{\ndocs(1-\volobs{j})}\,.
\end{equation}
It then makes sense to maximize the trace $\text{tr}(\Perm^T\mathbf{C})$.

Putting all together, the attack solves the problem
\begin{equation} \label{eq:graphm_obj}
	\Perm=\argmin_{\Perm\in\mathcal{P}}\quad (1-\alpha)\cdot||\coadv-\Perm\coobs\Perm^T||_F^2 - \alpha\cdot\text{tr}(\Perm^T\mathbf{C})\,,
\end{equation}
where $\alpha$ is the coefficient of the convex combination that the attacker must tune in order to optimize its performance.
Here, we have used $\mathcal{P}$ to denote the set of all valid column-selection permutation matrices $\Perm$.

The algorithms in the package\footnote{\url{http://projects.cbio.mines-paristech.fr/graphm/}} used by Pouliot et al.~\cite{pouliot2016shadow} to run this attack only work when the graphs have the same number of nodes, i.e., $\ntag=\nkw$, which is almost never the case in practice.
When $\ntag<\nkw$, by default the package fills the smallest graph with dummy nodes (e.g., it adds zeros to $\coobs$).
We show in Section~\ref{sec:eval} that this hampers the performance of the attack when $\ntag\ll\nkw$.

\subsection{Frequency Attack}

We explain the basic frequency attack by Liu et al.~\cite{liu2014search}.
In this attack, the adversary builds the frequency matrix for the tags $\Freqobs$, and uses the frequency matrix for keywords $\Freqadv$ as auxiliary-information.
The attacks assigns the keyword $\kw{i}$ to tag $\tagg{j}$ as
\begin{equation}
	\perm{j}=\argmin_{i\in[\nkw]} ||\fvecobs{j}-\fvecadv{i}||_2\,,
\end{equation}
where $||\cdot||_2$ the Euclidean norm for vectors.
The attack simply chooses, for each tag $\tagg{j}$, the keyword $\kw{i}$ whose frequency trend ($\fvecadv{i}$) is closest in Euclidean distance to the trend information of the tag ($\fvecobs{j}$).
This decision is independent for each tag, so several tags can be mapped to the same keyword (i.e., $\perm{\cdot}$ is not injective).

Liu et al.~also propose a more complex attack for a different query model where the client has preference for querying for keywords of a certain semantic category, and the adversary does not know this category a-priori.
We do not consider this setting in our work, for generality.
\section{Search and Access Pattern-Based Query Recovery Attack}
\label{sec:att}

We develop a query recovery attack that combines ideas from previous works~\cite{pouliot2016shadow,liu2014search}, but follows a pure Maximum Likelihood Estimation (MLE) approach and is orders of magnitude faster than the graph matching attack~\cite{pouliot2016shadow}.
In particular, we look for the mapping $\Perm$ that maximizes the likelihood of observing $\Volobs$, $\Freqobs$, $\Nqr$ and $\ndocs$ given the auxiliary information $\Voladv$ and $\Freqadv$.
We deliberately decide not to use the co-occurrence matrices $\coobs$ and $\coadv$ to help us estimate $\Perm$, for two reasons.
First, certain SSE techniques already hide keyword co-occurrence information~\cite{demertzis2020seal,patel2019mitigating}, as Blackstone et al.~\cite{blackstone2020revisiting} explain.
Second, it might be hard to obtain auxiliary keyword co-occurrence information $\coadv$ that is close to the actual data co-occurrence $\coobs$.
Our attack only uses background information from keyword volume $\Voladv$ and frequencies $\Freqadv$, which in many use cases can be easily obtained (e.g., from statistics about English word usage).

Formally, our attack solves the maximum likelihood problem
\begin{equation} \label{eq:bigproblem}
	\Perm=\argmax_{\Perm\in\mathcal{P}}\Pr(\Freqobs, \Nqr, \Volobs, \ndocs|\Freqadv, \Voladv, \Perm)\,.
\end{equation}

Note that it is not possible to exactly characterize this probability in practice.
Instead, we rely on a \emph{mathematical model} to characterize it.
We emphasize that there is no ``correct model'' for this task, but models that are close to the actual semantic properties of the database and the client's querying behavior will yield more accurate estimates of the true $\Perm$, while very unrealistic models will produce estimates with poor accuracy.
We use this mathematical model to derive our attack, and evaluate the performance of our attack with real data in Section~\ref{sec:eval}.

\subsection{Modeling the Observations}

We aim at characterizing $\Freqobs$, $\Nqr$, $\Volobs$, and $\ndocs$ given $\Freqadv$, $\Voladv$, and an assignment of tags to keywords $\Perm$.
We assume that the client's querying behavior and the response volumes are independent, i.e.,
\begin{equation} \label{eq:bigprob}
	\Pr(\Freqobs, \Nqr, \Volobs, \ndocs|\Freqadv, \Voladv, \Perm)=\Pr(\Freqobs, \Nqr|\Freqadv, \Perm)\cdot \Pr(\Volobs, \ndocs|\Voladv, \Perm)
\end{equation}

In our model, the number of queries the client makes in each time interval, $\Nqr$, follows an arbitrary distribution (independent of $\Perm$) that we represent as $\Pr(\Nqr)$.
The client chooses the keyword of each query \emph{independently} from other queries following the query frequencies $\Freqadv$.
This means that the number of queries for each keyword $i\in[\nkw]$ in time interval $k\in[\rho]$ follows a \emph{Multinomial distribution} with $\nqr{k}$ trials and probabilities given by $\fvecadv{k}$.
Formally,
\begin{align}
	\Pr(\Freqobs, \Nqr|\Freqadv, \Perm)&=\Pr(\Nqr)\cdot \Pr(\Freqobs|\Freqadv, \Nqr, \Perm)\\
			&=\Pr(\Nqr)\cdot \prod_{k=1}^\rho \Pr(\fvecobs{k}|\fvecadv{k}, \nqr{k}, \Perm)\\
			&=\Pr(\Nqr)\cdot \prod_{k=1}^\rho \nqr{k}!
			\prod_{j=1}^{\ntag}\frac{(\freqadv{\perm{j}}{k})^{\nqr{k}\freqobs{j}{k}}}{(\nqr{k}\freqobs{j}{k})!} \label{eq:probfreq}\,.
\end{align}

In our model, the number of documents in the encrypted database, $\ndocs$, is independent of $\Perm$, and the keywords of each encrypted document are chosen independently.
More precisely, given the relative volumes of the keywords from the auxiliary information $\Voladv=[\voladv{1},\dots,\voladv{\nkw}]$, each document has keyword $i\in[\nkw]$ with probability $\voladv{i}$.
This implies that the response volume when the client queries for $\kw{i}$ will be a Binomial random variable with $\ndocs$ trials and probability $\voladv{i}$, as in \eqref{eq:volbino}.
Formally,
\begin{align}
	\Pr(\Volobs, \ndocs|\Voladv, \Perm)&=\Pr(\ndocs)\cdot \Pr(\Volobs|\Voladv, \ndocs, \Perm)\\
			&=\Pr(\ndocs)\cdot \prod_{j=1}^{\ntag} 
			{\ndocs\choose {\ndocs\volobs{j}}} \voladv{\perm{j}}^{\ndocs\volobs{j}}(1-\voladv{\perm{j}})^{\ndocs(1-\volobs{j})}\,.\label{eq:probvol}
\end{align}

\subsection{Maximum Likelihood Estimator}

We use this model to find the $\Perm$ that maximizes $\Pr(\Freqobs, \Nqr, \Volobs, \ndocs|\Freqadv, \Voladv, \Perm)$.
We choose to maximize the logarithm of this probability instead to avoid precision issues (the problems are equivalent).
We can ignore the additive terms in the objective function that are independent of $\Perm$, since they do not affect the optimization problem.
The logarithm of equation \eqref{eq:bigprob} consists of two summands.
The first one is the logarithm of \eqref{eq:probfreq}.
The only term that depends on $\Perm$ here is 
\begin{equation} \label{eq:firstsum}
		\sum_{k=1}^\rho \sum_{j=1}^{\ntag}\nqr{k}\freqobs{j}{k}\cdot\log(\freqadv{\perm{j}}{k})\,.
\end{equation}

The second term of \eqref{eq:bigprob} is \eqref{eq:probvol}.
We can disregard $\Pr(\ndocs)$ and $\prod_{j=1}^{\ntag} {\ndocs\choose {\ndocs\volobs{j}}}$ since they do not depend on $\Perm$, and the remainder is:
\begin{equation} \label{eq:secondsum}
	\sum_{j=1}^{\ntag} \left[\ndocs\volobs{j} \log\voladv{\perm{j}} +\ndocs(1-\volobs{j}) \log(1-\voladv{\perm{j}})\right]
\end{equation}

We can write the problem of maximizing the summation of \eqref{eq:firstsum} and \eqref{eq:secondsum} in matricial form as follows.
First, we define two $\nkw\times\ntag$ cost matrices $\mathbf{C}_f$ and $\mathbf{C}_v$ whose $(i,j)$th entries are
\begin{align}
(\mathbf{C}_f)_{i,j} &\doteq -\sum_{k=1}^\rho \nqr{k}\freqobs{j}{k}\cdot\log(\freqadv{i}{k})\,,\label{eq:Cf}\\ 
(\mathbf{C}_v)_{i,j} &\doteq -\left[\ndocs\cdot\volobs{j}\cdot \log\voladv{i} +\ndocs(1-\volobs{j})\cdot \log(1-\voladv{i})\right]\,.\label{eq:Cv}
\end{align}
We add a negative sign to these matrices so that we can formulate the maximization problem in \eqref{eq:bigprob} as an \emph{unbalanced assignment problem}:
\begin{equation}\label{eq:hungproblem}
 \Perm=\argmin_{\Perm\in\mathcal{P}} \text{tr}(\Perm^T(\mathbf{C}_v+\mathbf{C}_f))\,.
\end{equation}

This problem can be efficiently solved with the Hungarian algorithm~\cite{kuhn1955hungarian}, whose complexity in the unbalanced case can be reduced to $O(\nkw\cdot\ntag + \ntag^2\cdot\log\ntag)$ as reported in~\cite{fredman1987fibonacci}.

\paragraph{Weighted Estimation.}

Sometimes, the adversary knows that their auxiliary volume information is more reliable than their frequency information, or vice-versa.
In these cases, it might make sense to assign more weight to their relative contribution to the optimization problem in~\eqref{eq:hungproblem}.
The adversary can do this by considering a combination coefficient $\alpha\in[0,1]$ and define the objective function as 
\begin{equation}\label{eq:hungproblem_alpha}
 \Perm=\argmin_{\Perm\in\mathcal{P}} \text{tr}(\Perm^T[(1-\alpha)\mathbf{C}_v+\alpha\mathbf{C}_f])\,.
\end{equation}

\section{Adapting the Attack against Privacy-Preserving SSE Schemes}
\label{sec:defenses}

So far, we have considered a generic SSE scheme that does not hide the access and query patterns.
This allows the adversary to compute the actual volume and frequency information, and carry out an attack with high accuracy (if the auxiliary information is accurate).
While there are no efficient techniques to hide the search patterns, there are many proposals that obfuscate the access patterns and/or response volumes.
In order to correctly assess the protection of these defenses, it is important to consider an attack performed by an adversary that is aware of the defenses implemented by the client.

In this section, we explain how to modify our attack to target particular privacy-preserving SSE schemes.
We adapt the attack by characterizing the probability of each keyword response volume given the auxiliary information, $\Pr(\Volobs|\Voladv, \ndocs, \Perm)$, when the defense takes place.
Following, we adapt the attack to three known privacy-preserving SSE schemes~\cite{chen2018differentially, patel2019mitigating, demertzis2020seal} that (partially) hide the access patterns, but our methodology applies to other existing (and future) defenses.
We introduce only the minimum information about these defenses required to understand how to adapt our attack against them, and refer to their papers for more details.
In Section~\ref{sec:discussion} we briefly discuss how to use our attack when the SSE scheme also hides search patterns.

\subsection{Differentially Private Access Patterns (CLRZ)}

The SSE scheme by Chen et al.~\cite{chen2018differentially} (that we denote CLRZ) hides the access patterns by adding random false positives and false negatives to the inverted index of the database.
This provides a certain level of indistinguishability between access patterns that can be expressed in terms of the differential privacy framework~\cite{dwork2008differential}.
Let $\TPR$ and $\FPR$ be the true positive and false positives rates of the defense, respectively.
First, the client generates an inverted index, i.e., a $\ndocs\times\nkw$ binary matrix whose $(\ell,i)$th element is 1 if the $\ell$th document has keyword $\kw{i}$, and 0 otherwise.
Then, each 0 in that matrix is flipped into a 1 with probability $\FPR$, and each 1 is set to 0 with probability $1-\TPR$.
This obfuscated matrix is used to generate the search index and determines which documents match each query.

Therefore, a document will match keyword $\kw{i}$ if this keyword was in the index before the obfuscation (probability $\voladv{i}$) and the defense didn't remove it ($\TPR$) or if the keyword was not in the original index $(1-\voladv{i})$, but the defense added it ($\FPR$).
This means that, after applying the defense, the probability that a document has keyword $i$ is
\begin{equation} \label{eq:newprob}
 \voladv{i}\cdot\TPR + (1-\voladv{i})\cdot\FPR\,.
\end{equation}

We can adapt the attack against this defense by replacing $\voladv{i}$ in \eqref{eq:Cv} by \eqref{eq:newprob}.

\subsection{Differentially Private Volume (PPYY)}

The defense by Patel et al.~\cite{patel2019mitigating} (that we denote PPYY) assumes that the server stores independent document and keyword pairs (i.e., the server stores a copy of each document for each keyword this document has).
The documents are stored in a hash table such that $H(\kw{i}||k)$ points to the $k$th document that has keyword $\kw{i}$, or to any random document if there are less than $k$ documents with keyword $\kw{i}$.
When querying for keyword $\kw{i}$, the client sends the hashes $H(\kw{i}||1), H(\kw{i}||2), \dots, H(\kw{i}||v)$ (for a certain volume $v$) and receives the documents in those positions of the hash table.
Since the server is storing independent document-keyword pairs, queries for different keywords are completely uncorrelated and thus it is not possible to infer information from the access pattern structure (such as the co-occurrence matrix $\coobs$).
However, the scheme must use a different volume for each keyword, since padding each keyword to the same volume is overly expensive.

Patel et al.~propose to obfuscate the volume by adding Laplacian noise to it, plus a constant value to ensure that this extra volume is never negative.
If the Laplacian noise plus constant is negative for a keyword, the scheme would be lossy, i.e., there would be false negatives when querying for that keyword.

Let $\epsilon$ be the privacy parameter of the scheme.
Adding Laplacian noise with scale $2/\epsilon$ ensures $\epsilon$-differential privacy for the leaked volumes, i.e., for low values of $\epsilon$ (e.g., $\epsilon<1$) an adversary would not be able to distinguish between two keywords whose response volumes differ by a single document.

In order to ensure a negligible probability that Laplacian noise plus a constant is negative for any keyword, we follow the approach by Patel et al.~\cite{patel2019mitigating}:
The probability that at least one of $\nkw$ independent samples from $\text{Lap}(2/\epsilon)$ is smaller than a constant $2t/\epsilon$ is upper bounded by $\nkw\cdot e^{-t}$.
We want this probability to be negligible, so we set $\nkw\cdot e^{-t}=2^{-64}$ and find that $t=\log n+64\cdot\log 2$.

Therefore, if we use $\voltrue{j}$ to denote the \emph{true volume} of keyword $\kw{\perm{j}}$, and $\lceil\cdot\rceil$ denotes the ceiling function, the observed volume for tag $\tagg{j}$ would be
\begin{equation} \label{eq:ppyy}
  \volobs{j} = \voltrue{j} + \lceil\text{Lap}(2/\epsilon) + 2(\log \nkw + 64 \cdot \log 2)/\epsilon\rceil\,.
\end{equation}
We use the ceiling function since volumes need to be integers.
Note that the overhead of this scheme increases with the number of keywords $\nkw$, because the constant padding term needs to ensure that none of the keywords gets negative padding.

We use this expression directly to compute $\Pr(\Volobs|\Voladv, \ndocs, \Perm)$.
In this case, we cannot derive a closed-form expression for $\mathbf{C}_v$ and compute it as follows: for each $i\in[\nkw]$, compute the convolution between the probability mass functions of $\text{Bino}(\ndocs, \voladv{i})$ and $\text{Lap}(2/\epsilon)$ shifted by constant $2(\log \nkw + 64 \cdot \log 2)/\epsilon$ and discretized with the ceiling function.
Then, $(\mathbf{C}_v)_{i,j}$ is the value of the resulting function evaluated at $\volobs{j}$.

\subsection{Multiplicative Volume Padding (SEAL)}
\label{sec:seal}

The SEAL defense technique, proposed by Demertzis et al.~\cite{demertzis2020seal}, has two parameters, $\alpha$ and $x$.
In SEAL, the server stores the database in $2^\alpha$ ORAM blocks, so that it is not possible to tell which document within each block is accessed each time.
This means that SEAL leaks \emph{quantized} versions of the true access patterns.
Additionally, SEAL pads the response volume of each query to the closest power of $x$.

Our attack uses the access patterns to identify whether or not two queries are distinct (i.e., to infer the search pattern).
We note that it is possible to obfuscate the search pattern by choosing a small enough $\alpha$ to cause collisions in the quantized access patterns of different queries.
However, we argue that this requires such a small value of $\alpha$ that might significantly affect the efficiency of SEAL, so we still consider that queries for distinct keywords generate distinct access patterns, and thus SEAL leaks the search pattern.
Note that this is the case in the original work~\cite{demertzis2020seal}, since the authors use large values of $\alpha$ (that are close to $\log_2 \ndocs$).

Let $\voltrue{j}$ be the true volume of keyword $\kw{\perm{j}}$ in the dataset.
The observed volume when querying for this keyword in SEAL is $x^{\lceil \log_x\voltrue{j}\rceil}$.
We compute $\mathbf{C}_v$ as follows: for each $i\in[\nkw]$, compute the probability that $\text{Bino}(\ndocs, \voladv{i})$ falls between each interval $(x^{k-1}, x^{k}]$ for $k\in[\lceil\log_x\ndocs\rceil]$.
Denote this probability by $\text{Prob}(k, i)$.
Then, $(\mathbf{C}_v)_{i,j}$ is $\text{Prob}(\lceil \log_x\volobs{j}\rceil, i)$.

\section{Evaluation}
\label{sec:eval}

In this section, we compare the performance of our attack with the graph matching attack by Pouliot and Wright~\cite{pouliot2016shadow} and the frequency attack by Liu et al.~\cite{liu2014search}, and evaluate our attack against the three defenses we considered above~\cite{chen2018differentially, patel2019mitigating, demertzis2020seal}.
We denote our attack by $\attack$ (search and access pattern-based attack) to distinguish it from $\graphm$~\cite{pouliot2016shadow} and $\liu$~\cite{liu2014search}.

We use Python3.7 to implement our experiments\footnote{Our code is available at \url{https://github.com/simon-oya/USENIX21-sap-code}} and run then in a machine running Ubuntu 16.04 in 64-bit mode using 32 cores of an Intel(R) Xeon(R) CPU (2.00GHz) with 256 GB of RAM.
We use Scipy's implementation of the Hungarian algorithm to run our attack (i.e., to solve \eqref{eq:hungproblem}).

\paragraph{Experimental Setup.}

We use two publicly available email datasets to build the client's database and the server's auxiliary information.
The first dataset is Enron email corpus,\footnote{\url{https://www.cs.cmu.edu/~./enron/}} which contains $30\,109$ emails from Enron corporation, and is popular among related works~\cite{cash2015leakage, islam2012access, liu2014search, pouliot2016shadow, zhang2016all}.
The second dataset, used by Cash et al.~\cite{cash2015leakage}, is the \textit{java-user} mailing list from the lucene project.\footnote{\url{https://mail-archives.apache.org/mod_mbox/lucene-java-user/}}
We took the emails of this mailing list from September 2001 until May 2020 (around $66\,400$ emails).
Each email is one document in the dataset, and its keyword list is the set of words in the main body of the email that are part of an English dictionary, excluding English stopwords.
We use Python's NLTK corpus\footnote{\url{https://www.nltk.org/howto/corpus.html}} to get a list of all English words and stopwords.

We select the $3\,000$ most frequent keywords to build a set $\Delta_{3\,000}$ for each dataset.
Then, in each experiment run, given $\nkw$, we generate the keyword universe $\Delta$ by randomly selecting $\nkw$ keywords from $\Delta_{3\,000}$.
In each experiment run, we perform a random keyword selection and a random split of the dataset; we use half of the documents as the actual client's dataset, and give the other half to the adversary to use as auxiliary information to compute $\Voladv$ and $\coadv$.

We could not find any public database with actual user query information for either of the databases.
This is a common problem when evaluating attacks that use query frequency, as observed by Liu et al.~\cite{liu2014search}.
Therefore, we use query information from Google Trends\footnote{\url{https://trends.google.com/trends}} to generate client queries~\cite{liu2014search}.
For each keyword in $\Delta_{3\,000}$, we get its search popularity for the past 260 weeks (ending in the third week of May 2020).
We store these popularity values in a $3\,000\times 260$ matrix.
In each experiment run, given a particular keyword universe $\Delta$ of size $\nkw$, we take the popularity of each of those keywords in the last 50 weeks and store it in a $\nkw\times 50$ matrix that we denote $\Freqtrue$.
Then, we normalize the columns of this matrix so that they add up to one.
The observation time is always 50 weeks, and we vary the \emph{average} number of queries per week ($\nqravg$) that the client performs.
We generate the actual number of queries that the client performs for keyword $\kw{i}$ in week $k$ by sampling from a Poisson distribution with rate $\nqravg\cdot \freqobs{i}{k}$, where $\freqtrue{i}{k}$ is the $(i,k)$th element of $\Freqtrue$.

Since giving the true frequency information to the adversary would be unrealistic, we give the adversary outdated frequency information instead.
For a certain week offset $\tau$, the adversary's auxiliary frequency information is $\freqadv{i}{k}=\freqtrue{i}{k-\tau}$.
Note that the observed frequencies $\freqobs{j}{k}$ will only approach $\freqtrue{i}{k}$ as $\nqravg\to\infty$.
In most of our experiments, we set a very low number of average queries per week ($\nqravg=5$), so the information the adversary gets from the query frequencies is very limited.
We think this approach is more realistic than giving the adversary frequencies perturbed with Gaussian noise~\cite{liu2014search}.

We perform 30 runs of each of our experiments (in parallel), using a different random seed for each.
This randomness affects the keyword selection, the dataset split, the query generation, and the defense obfuscation techniques.
The attacks are deterministic.
We measure the query recovery \emph{accuracy}, which we compute by counting how many of the client's queries the attack recovers correctly and normalizing by the total number of queries (with possibly repeated keywords).
For completeness, we also report the percentage of unique keywords recovered in each experiment in the Appendix.

\subsection{Preliminary Experiments for Our Attack}

\begin{figure*}[t]
	\begin{minipage}{0.49\linewidth}
		\centering
		\includegraphics[width=\linewidth]{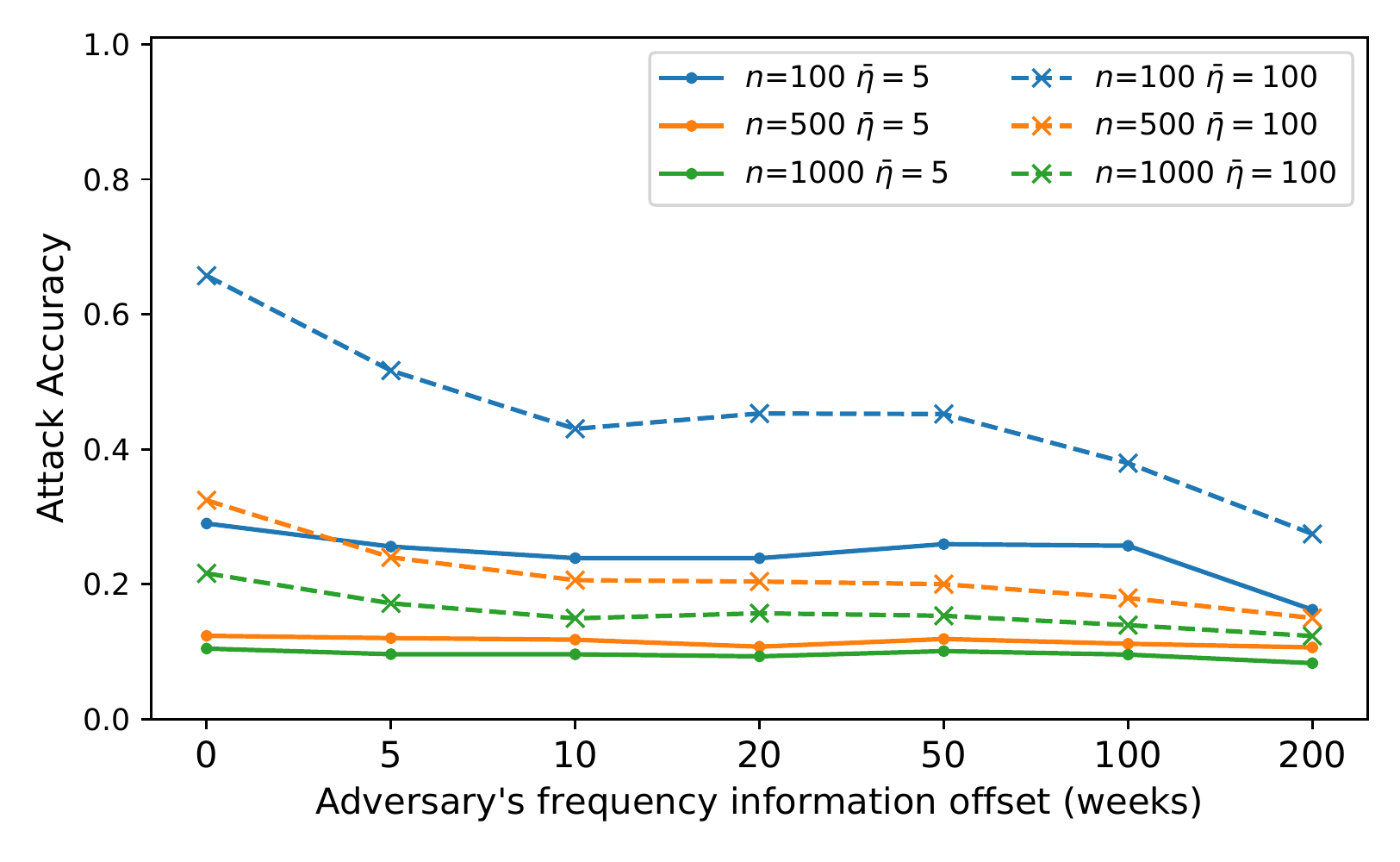}\\
		\caption{Effect of outdated frequency information in the performance of $\attack$ against a basic SSE in Enron dataset.}
		\label{fig:offset}
	\end{minipage} \hfill
	\begin{minipage}{0.48\linewidth}
		\centering
		\includegraphics[width=0.95\linewidth]{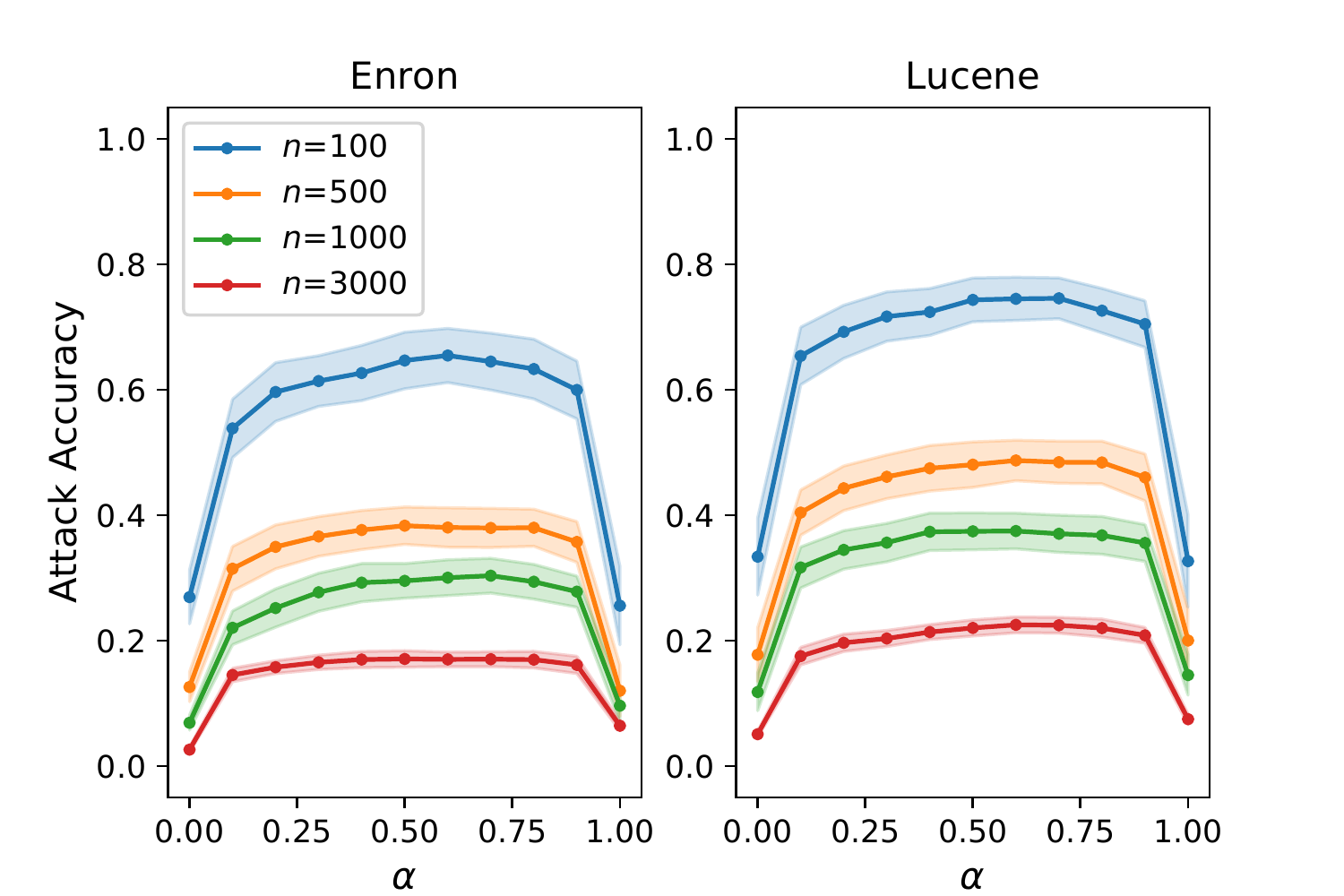}\\
		\caption{Effect of $\alpha$ in the performance of $\attack$ against a basic SSE ($\nqravg=5$ queries per week, 50 weeks).}
		\label{fig:hung_with_alpha}
	\end{minipage}
\end{figure*}

We perform a preliminary experiment to observe the effect of the auxiliary information offset $\tau$ in $\attack$.
We perform the attack on Enron dataset using only frequency information, i.e., $\alpha=1$ in \eqref{eq:hungproblem_alpha}, and show these results in Figure~\ref{fig:offset} for different sizes of the keyword universe $\nkw$ and average number of weekly queries $\nqravg$.
We see that the frequency information slowly degrades with the offset (we see a slight peak at 50 weeks when $\nkw=100$, since this is almost one year and some query behaviors repeat yearly).
Also, the accuracy decreases with the keyword universe size $\nkw$, since estimating the keyword of each query becomes harder when there are more possible keywords to choose from.
We use an offset of $\tau=5$ in the remainder of the evaluation, since most of our experiments are for $\nqravg=5$ and we see that the accuracy degradation stabilizes after that.

We carry out a second experiment to understand how $\attack$ benefits from both access and search pattern leakage.
We set $\nqravg=5$ (average of 250 queries in total over 50 weeks) and vary $\alpha\in[0,1]$.
We show the attack's accuracy for different keyword universe sizes $\nkw$ in Figure~\ref{fig:hung_with_alpha}.
The lines are the average accuracy of the attacks, and the shades represent the 95\% confidence interval.
The results are qualitatively similar in both datasets, although it is slightly easier to identify keywords in Lucene.
This experiment reveals that using either volume ($\alpha=0$) or frequency ($\alpha=1$) information alone provides low accuracy values (e.g., below $15\%$ for $\nkw=1\,000$ in Enron).
However, combining both types of information provides an outstanding boost (the accuracy is more than twice as large than when using either type of information by itself).
In the remaining experiments, we use the pure maximum likelihood estimator ($\alpha=0.5$) configuration for $\attack$.

\subsection{Comparison with Other Attacks}

We compare the performance of $\attack$ with the graph matching attack by Pouliot et al.~\cite{pouliot2016shadow} ($\graphm$) and the frequency attack by Liu et al.~\cite{liu2014search} ($\liu$).
We use the GraphM package\footnote{\url{http://projects.cbio.mines-paristech.fr/graphm/}} to solve the graph matching problem of $\graphm$.
This package offers different graph matching algorithms, and we use the PATH algorithm~\cite{zaslavskiy2008path}, since it provides the best results~\cite{pouliot2016shadow}.

\begin{figure}[t]
	\centering
	\includegraphics[width=0.95\columnwidth]{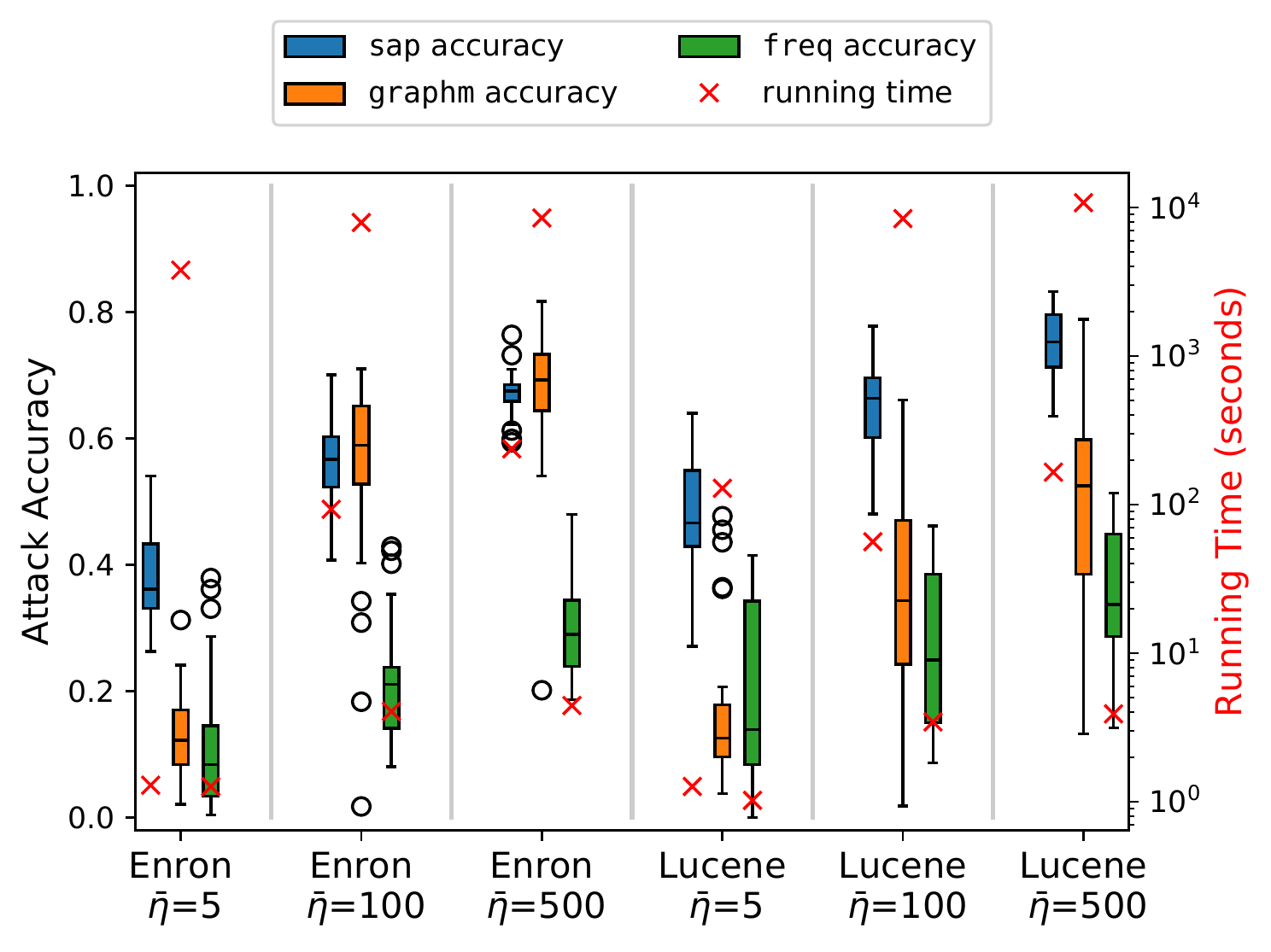}
	\caption{Comparison of the query recovery accuracy (boxes) and running time (\textcolor{red}{$\times$}) of attacks in different datasets with $\nqravg$ queries per week (50 weeks), with $\nkw=500$ keywords.}
\label{fig:attcomp_box}
\end{figure}

We show the results of our experiments in Figure~\ref{fig:attcomp_box}. 
The boxes show the accuracy of the attacks (left axis), and the red crosses (\textcolor{red}{$\times$}) represent their average running time (right axis, logarithmic).
We use the pure MLE approach for $\attack$ ($\alpha=0.5$) and plot the results of $\graphm$ with the best performing $\alpha$ each time (we tried $\alpha=0$ to $\alpha=1$ with steps of $0.1$).
We use $\nkw=500$ for this plot (we do not use a larger number since the running times of $\graphm$ become unfeasible).

Our attack ($\attack$) is approximately four times more accurate than $\graphm$ and $\liu$ when the client performs few queries ($\nqravg=5$) in both datasets.
The performance of all the attacks increase as the adversary observes more queries, but $\attack$ takes the lead in most cases.
For $\nqravg=500$ (a total of $\approx 25\,000$ queries observed), in Enron dataset, $\graphm$ achieves a slightly higher average accuracy than $\attack$.
However, note that the running time of $\graphm$ is always approximately two orders of magnitude larger than $\attack$ (note the logarithmic right axis).

Our experiments reveal that $\graphm$ heavily relies on observing almost all possible keywords to achieve high query recovery rates.
We argue that this is a consequence of how the graph matching problem \eqref{eq:graphm_obj} is framed.
Note that, when $\ntag\ll\nkw$, the matrix $\Perm\coobs\Perm^T$ will have many zero entries (the solver actually fills the smallest graph with dummy nodes, as we explain in Section~\ref{sec:graphm}).
In this case, a good strategy to minimize \eqref{eq:graphm_obj} is to simply choose the permutation $\Perm$ that cancels the largest terms in $\coadv$.
This permutation is not necessarily a good estimate of the the correct assignment of tags to keywords.
This could potentially be solved by shrinking $\coadv$ instead, i.e., $||\Perm^T\coadv\Perm-\coobs||_F^2$ and/or using a norm that does not give more weight to large terms (e.g., opting for an $L1$-norm instead of the Frobenius or $L2$-norm).
We note that improving this attack might still be unprofitable, since keyword co-occurrence is completely infective against recent SSE schemes~\cite{blackstone2020revisiting}.

In conclusion, the experiments confirm the relevance of our attack, since 1) it is computationally efficient, 2) it outperforms $\liu$, 3) it outperforms $\graphm$ when the client does not query for all possible keywords, which we argue is a realistic scenario.
Also, our attack does not require background knowledge of keyword co-occurrence and is easily adaptable against defenses.
This adaptability is key towards assessing the effectiveness of these defenses, as we show next.

\subsection{Performance of $\attack$ against Defenses}

We evaluate the performance of $\attack$ against the three defenses we considered in Section~\ref{sec:defenses}.
We give the adversary the frequency information with an offset of $\tau=5$ weeks and we set the observation time to 50 weeks, as before.
The average number of queries per week is $\nqravg=5$ (i.e., average of $250$ queries in total).
We use this arguably low number to show that, even with a small number of queries, frequency information can really help the adversary.
Again, we consider the pure MLE approach of $\attack$ \eqref{eq:hungproblem}, i.e., $\alpha=0.5$.
We evaluate the performance of the attack with up to $\nkw=3\,000$, since it is computationally efficient. 

\begin{figure*}[t]
\begin{minipage}{\linewidth}
	\begin{minipage}{0.49\linewidth}
		\centering
		\includegraphics[width=0.95\linewidth]{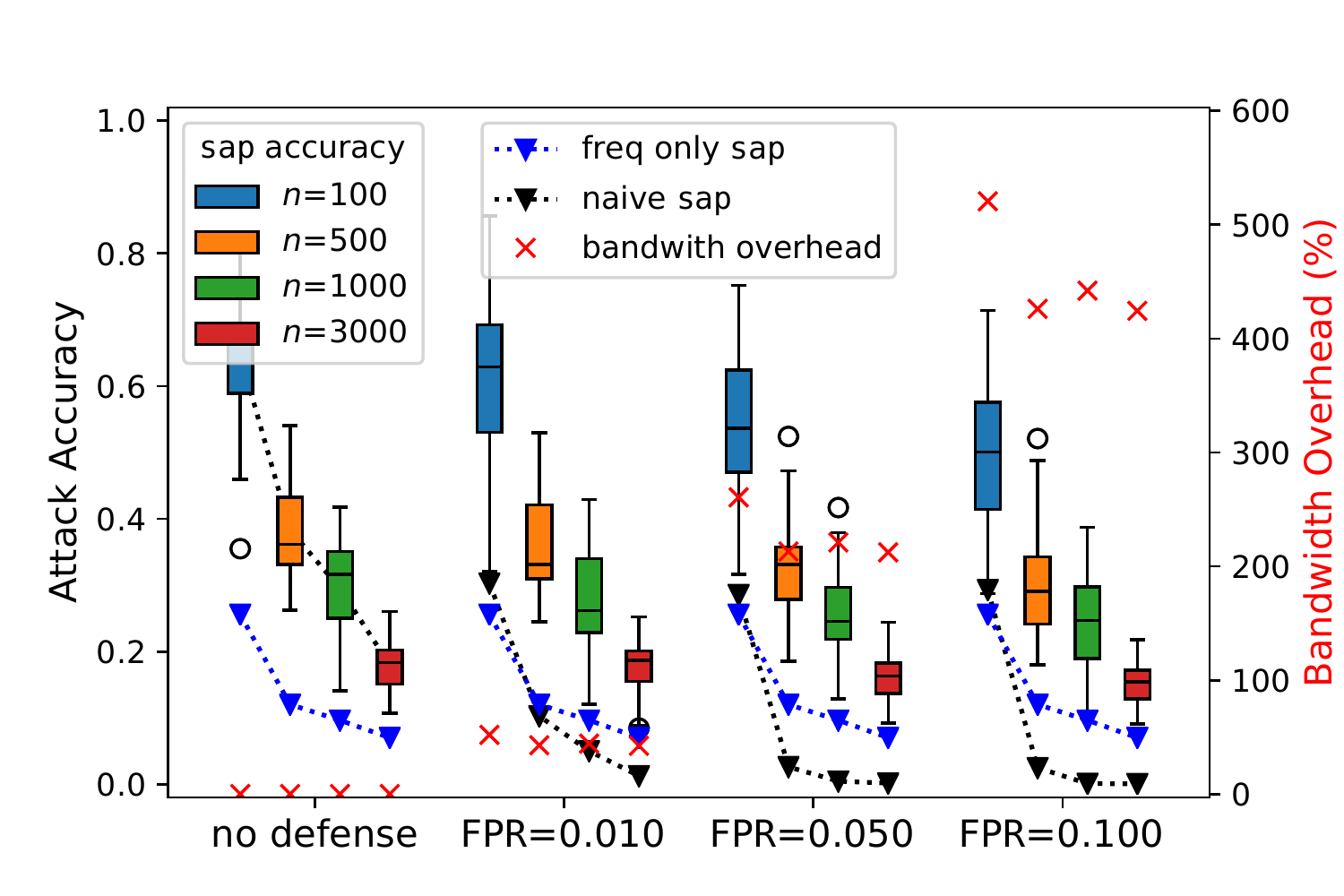}\\
		(a) Enron dataset
	\end{minipage} \hfill
	\begin{minipage}{0.49\linewidth}
		\centering
		\includegraphics[width=0.95\linewidth]{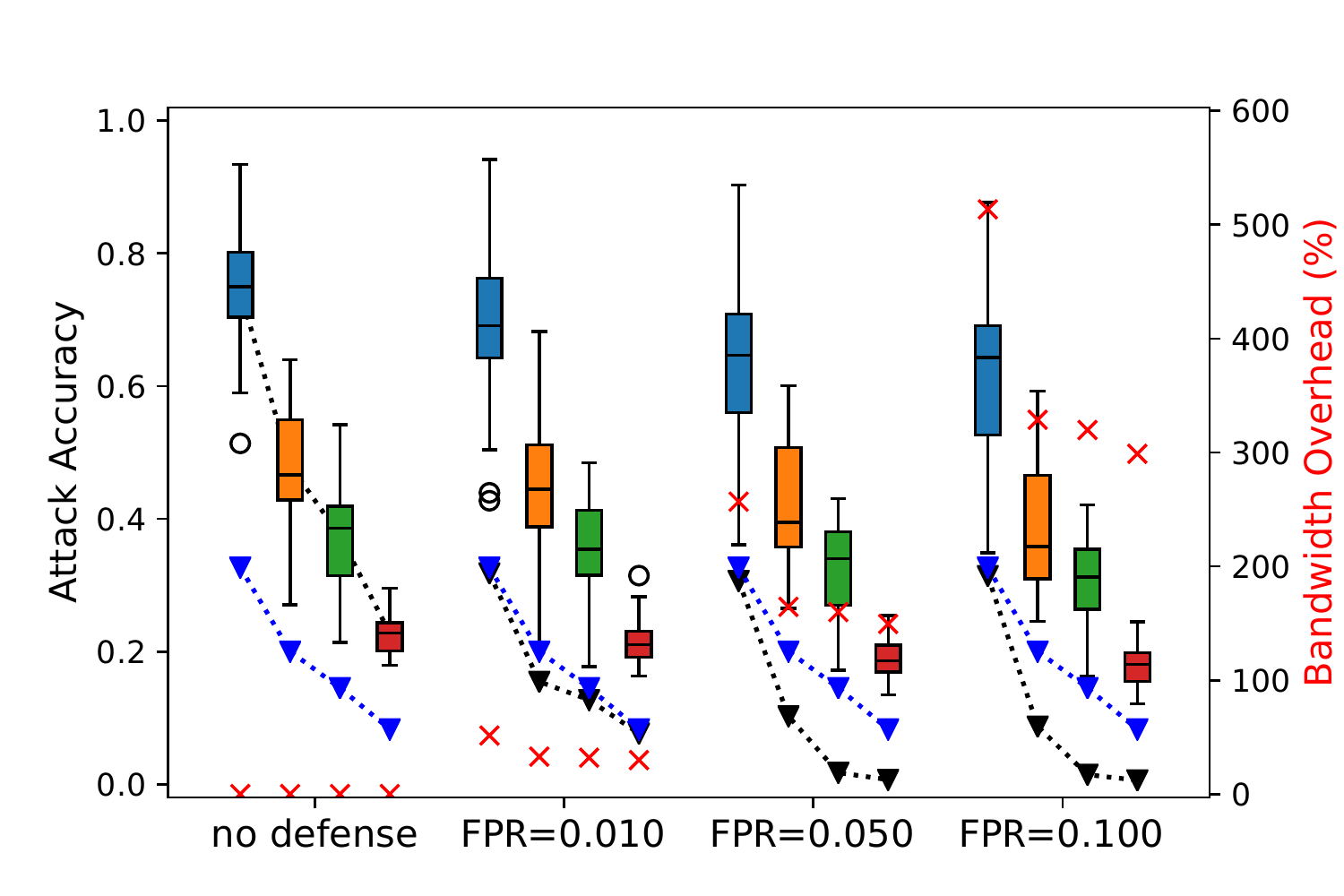}\\
		(b) Lucene dataset
	\end{minipage}
\end{minipage}
\caption{Accuracy of $\attack$ against CLRZ defense configured with $\TPR=0.999$ and varying $\FPR$ (50 weeks, $\nqravg=5$ queries/week).}
\label{fig:vs_clrz}
\end{figure*}

\paragraph{Performance against CLRZ~\cite{chen2018differentially}.}
We set the true positive rate of CLRZ to $\TPR=0.999$ and vary the $\FPR$ between $0.01$, $0.05$, and $0.1$. 
Figure~\ref{fig:vs_clrz} shows the results in Enron (a) and Lucene (b).
We generate the boxes using the accuracy values of $\attack$ in 30 runs of the experiment.
The dotted black lines represent the mean accuracy of $\attack$ without adapting it against this defense, i.e., this would be the performance if the adversary was \emph{unaware} of the defense.
As a reference, the dotted blue lines show the performance of $\attack$ using frequency information only ($\alpha=1$).
The red crosses (\textcolor{red}{$\times$}) represent the bandwidth overhead of the defense (marked in the right axis), that we compute as follows.
Let $N_R$ be the total number of documents returned by the server in a run of the experiment, and let $N_r$ be the number of documents that would be returned if the defense had not been applied.
Then, the overhead percentage is $(N_R/N_r-1)\cdot 100$.
This value is only a reference, since the actual overhead depends on implementation details.

Increasing $\FPR$ improves the protection of the defense.
For example, with $\nkw=1\,000$ keywords in Lucene, the attack accuracy drops from $37\%$ (no defense) to $\approx 1\%$ ($\FPR=0.1$) against the naive attack (black doted line).
However, by adapting the attack against the defense, the accuracy increases back to $30\%$.
We observe this behavior in both datasets and for all values of $\nkw$, which confirms that our attack is able to almost ignore the defense.
Note that the maximum $\FPR$ value we consider ($\FPR=0.1$) indicates that around $10\%$ of the whole dataset is returned in each query, which is already unrealistically high in real cases (the overhead is betwen $400\%$ and $500\%$ when $\FPR=0.1$).

\paragraph{Performance against PPYY~\cite{patel2019mitigating}.}

We configure PPYY with privacy values $\epsilon=1$, $0.2$, and $0.1$.
Note that smaller values of $\epsilon$ increase the amount of padding (and the overall privacy the scheme provides).
Typically, in differential privacy scenarios, values of $\epsilon<1$ are considered high privacy regimes.
Patel et al.~\cite{patel2019mitigating} use $\epsilon=0.2$ in their cost evaluation.

\begin{figure*}[t]
\begin{minipage}{\linewidth}
	\begin{minipage}{0.49\linewidth}
		\centering
		\includegraphics[width=0.95\linewidth]{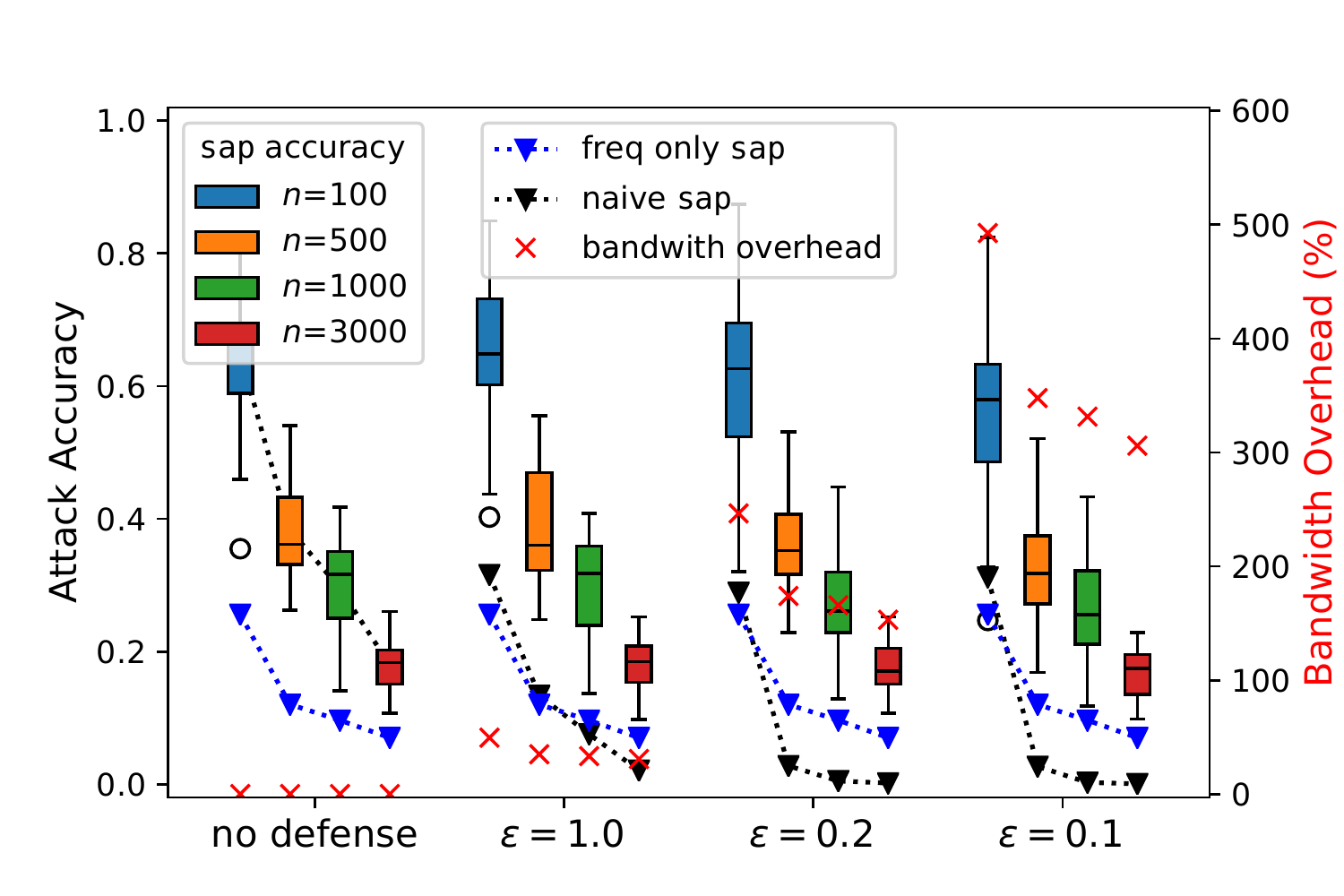}\\
		(a) Enron dataset
	\end{minipage} \hfill
	\begin{minipage}{0.49\linewidth}
		\centering
		\includegraphics[width=0.95\linewidth]{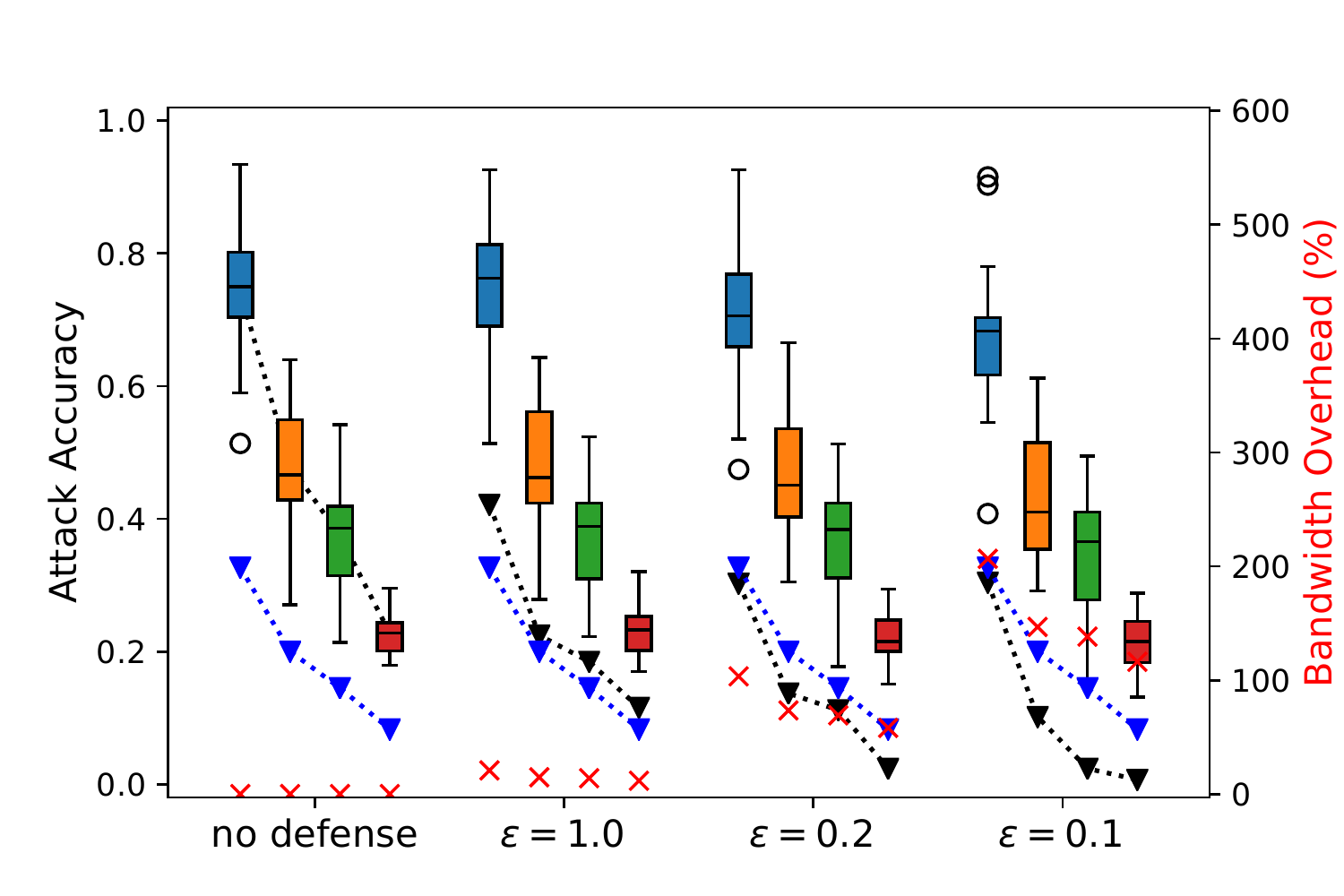}\\
		(b) Lucene dataset
	\end{minipage}
\end{minipage}
\caption{Accuracy of $\attack$ against PPYY defense with different privacy values $\epsilon$ (50 weeks, $\nqravg=5$ queries/week).}
\label{fig:vs_ppyy}
\end{figure*}

Figure~\ref{fig:vs_ppyy} shows the results in the same format as in the previous case.
When computing the bandwidth overhead, we only take into account the overhead caused by the extra padding as explained above.
The original scheme incurs extra overhead, e.g., due to the type of hashing technique used to store the database.
We refer to their paper for the detailed cost analysis of this defense.
Our goal with this experiment is to show the effectiveness of Laplacian noise as a volume-hiding technique.

The results are qualitatively (and quantitatively) very close to the results for the previous defense.
Values of $\epsilon=0.1$ seem to be effective at reducing the accuracy of the naive attack (dropping from $37\%$ accuracy to $\approx 2\%$ in Lucene with $\nkw=1\,000$) but, when tailoring the attack against the defense, it recovers queries with a similar accuracy as when no defense is applied ($35\%$ in the aforementioned case).

The reason for this is the following: even though $\epsilon=0.1$ is a high differential privacy regime, this privacy notion only ensures that queries for keywords whose response volume differs \emph{in one unit} are indistinguishable.
As Patel et al.~admit~\cite{patel2019mitigating}, in some settings this privacy definition might be unreasonable.
This seems to be the case for the datasets we consider, and more generally it seems unrealistic to consider an optimistic setting where the only queries the adversary wants to distinguish are for keywords whose response volume differs in one document.

\paragraph{Performance against SEAL~\cite{demertzis2020seal}.}

As we explain in Section~\ref{sec:seal}, we assume that there are no collisions between the quantized access patterns that SEAL leaks, so that the scheme implicitly reveals the search pattern and the adversary can compute the query frequencies of each tag.
We vary the multiplicative padding $x$ between $2$, $3$, and $4$.
Recall that SEAL pads the volume of each keyword to the next power of $x$, and thus the overhead percentage is always smaller than $(x-1)\cdot 100$.

\begin{figure*}[t]
\begin{minipage}{\linewidth}
	\begin{minipage}{0.49\linewidth}
		\centering
		\includegraphics[width=0.95\linewidth]{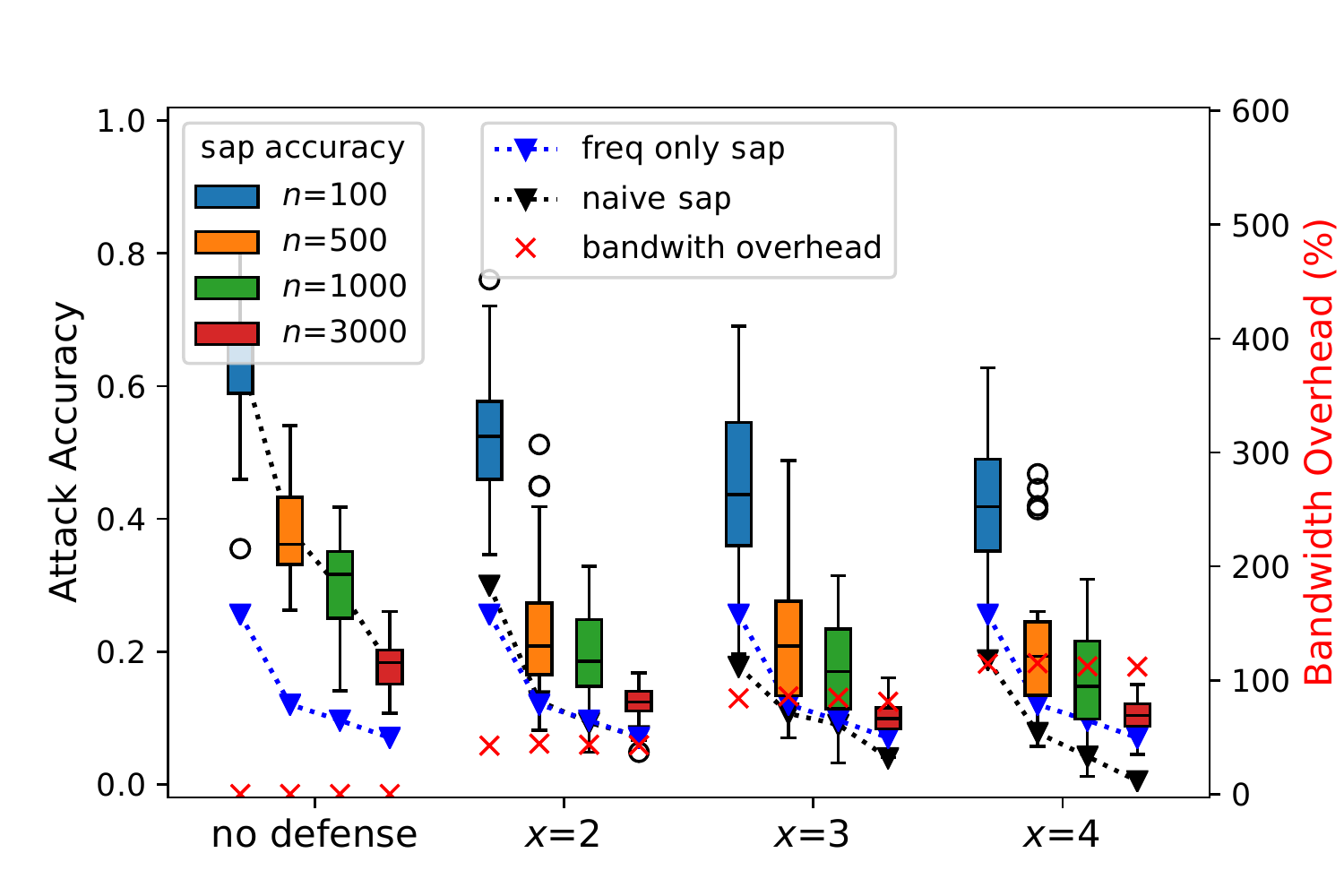}\\
		(a) Enron dataset
	\end{minipage} \hfill
	\begin{minipage}{0.49\linewidth}
		\centering
		\includegraphics[width=0.95\linewidth]{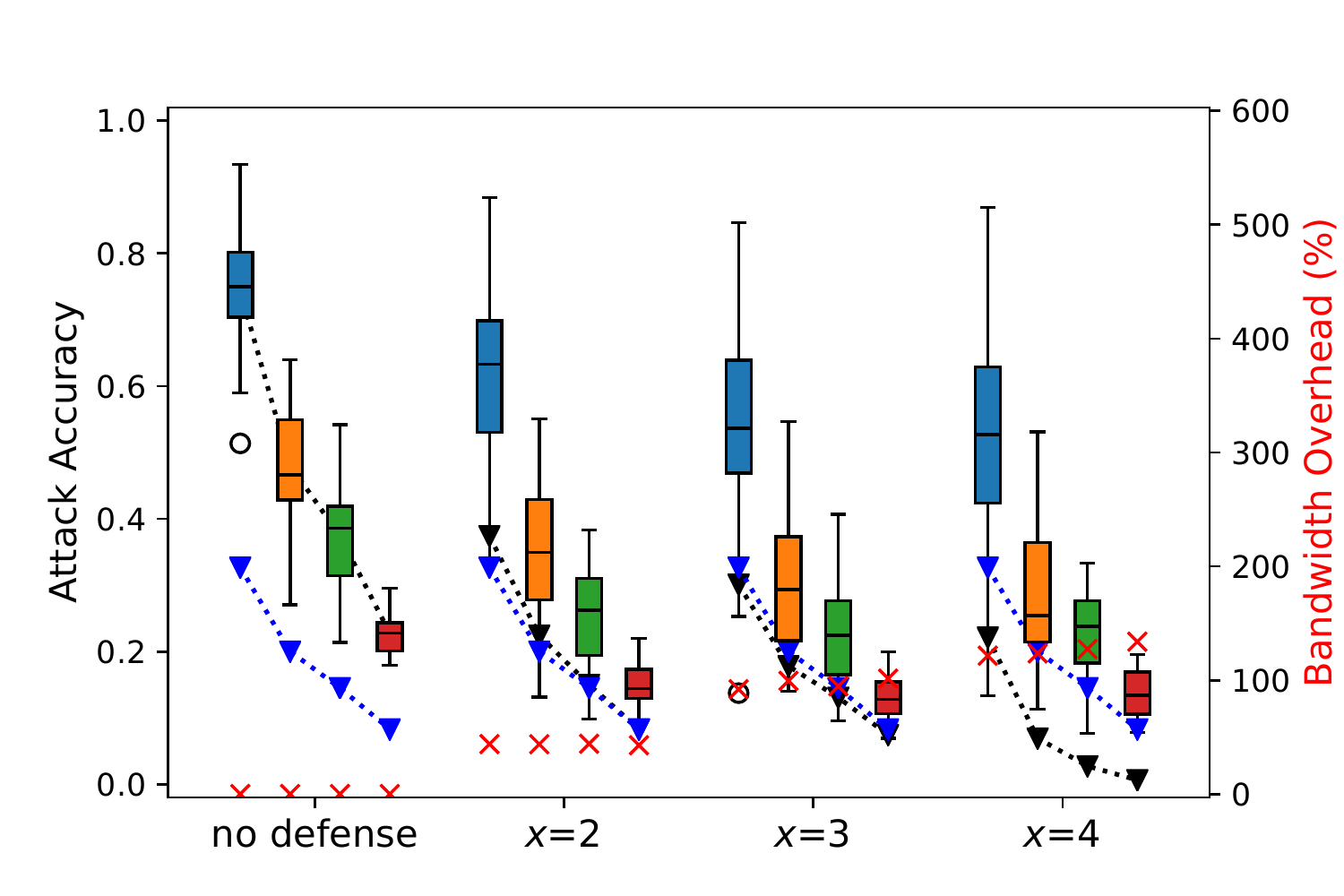}\\
		(b) Lucene dataset
	\end{minipage}
\end{minipage}
\caption{Accuracy of $\attack$ against SEAL defense for different values of multiplicative volume padding $x$ (50 weeks, $\nqravg=5$ queries/week).}
\label{fig:vs_seal}
\end{figure*}

Figure~\ref{fig:vs_seal} shows the results.
Following the example above (Lucene with $\nkw=1\,000$), the attack accuracy drops from $37\%$ to $3\%$ with a padding parameter $x=4$.
A defense-aware attacker brings the accuracy up to $23\%$, which is still a significant value, but below the performance of the attack against the other two defenses.
The results show that multiplicative volume padding is a highly efficient volume-hiding technique, since it achieves significantly more protection than the other two, with less bandwidth overhead.

We highlight that in all these experiments both the volume and the frequency information contribute the attack's success.
This can be seen in the figures by noting that the boxes are significantly above the dashed blue lines (frequency-only $\attack$).

\section{Discussion: Preventing Frequency Leakage}
\label{sec:discussion}

Throughout the paper, we have only considered defenses that obfuscate the access pattern and/or response volume.
Completely hiding the volume information would require returning the same number of documents in response to every query, which is unreasonable in terms of bandwidth overhead~\cite{kamara2018encrypted, demertzis2020seal}.
We have seen that, even when the volume is obfuscated, the frequency information (derived from the search pattern) surprisingly contributes to the success of our query identification attack.
This is true even when the user only performs 5 queries per week and the observation time is 50 weeks (even if we consider keyword universes of size $\nkw=3\,000$).
Below we discuss some  alternatives for hiding this frequency information which we believe is key towards achieving effective privacy-preserving SSE schemes.

\paragraph{Hiding the Search Pattern with Collisions.}
Hiding the search pattern implies that the adversary is not able to tell whether or not a query has been repeated.
This prevents the adversary from (correctly) assigning tags to queries and thus from computing observed query frequencies.

One option to hide the search pattern among groups of keywords is to create collisions between access patterns, i.e., force queries for different keywords to return the same set of documents.
This idea of ``merging keywords'' is similar to the Secure Index Matrix~\cite{islam2012access} and, to some extent, to the Group-Based Construction~\cite{liu2014search}.
In practice, it is still not clear how to provide privacy by grouping keywords while keeping the overhead of the scheme under reasonable bounds.
This is because it is more efficient to merge keywords that appear in a similar set of documents, but these keywords would very likely have a similar semantic meaning (e.g., medical terms will appear in similar documents).
Therefore, one might argue that, in this case, guessing that a keyword belongs to a group of words with similar semantic meaning can already be a privacy violation.

\paragraph{Hiding the Search Pattern with Fresh Randomness.}
The schemes we have considered in this work leak the search pattern because the same keyword always produces the same access pattern.
A scheme that generates access patterns with fresh randomness could prevent this from happening.
A possible solution for this would be using an ORAM (e.g., TwoRAM~\cite{garg2016tworam}) scheme to hide which documents are retrieved from the dataset, and randomize the volume padding independently in every query.
The problem with this solution is that ORAM-based SSE schemes incur considerable communication costs.

Even if the client was able to generate independent random access patterns for each query, the adversary could try to cluster similar access patterns together (two queries for the same keyword might still produce statistically similar access patterns since they aim to return the same set of documents).
This clustering algorithm would be used to tag the observed queries.
This tagging process would have some errors, that in the end would lower the accuracy of the query identification attack.
It is however unclear how to build an \emph{efficient} SSE scheme with independent access pattern obfuscation for each query such that access patterns are hard to cluster by keyword.

\paragraph{Hiding the Query Frequencies with Dummy Queries.}
A third alternative that has not been thoroughly explored in the literature is, instead of hiding the search patterns, obfuscating the query frequencies themselves by performing \emph{dummy queries}.
There are two immediate problems with this approach: first, it is not clear how to choose \emph{when} to generate dummy queries without leaking whether the query is real or not through timing information.
Generating a deterministic set of dummy queries for each real query~\cite{liu2014search} reveals more information and is less efficient than just merging these keywords in the search index (the first solution we mentioned in this section).
A possible solution to this problem could come from anonymous communication technologies that already use traffic analysis-resistant dummy strategies (e.g., the Poisson cover traffic in Loopix~\cite{piotrowska2017loopix}).
Another problem of hiding query frequencies with dummy queries is \emph{how} to choose the keywords of the dummy queries without requiring the client to store the set of all possible keywords in its local storage.

Even if the client implemented a dummy generation strategy, the adversary would know the particulars of this method and could adapt the attack accordingly, making corrections to the observed frequencies and limiting the effectiveness of the defense.
Therefore, hiding the true frequency of queries with reasonable bandwidth overhead might be challenging.

\section{Conclusions}
\label{sec:conclusions}

In this work, we propose a query recovery attack against privacy-preserving Symmetric Searchable Encryption (SSE) schemes that support point queries.
We derive this attack by setting up a maximum likelihood estimation problem and computing its solution by solving an unbalanced assignment problem.
Unlike previous attacks, our proposal combines both volume information, computed from the access pattern leakage, and frequency information, obtained from the search pattern leakage.
We show that, even in cases where taking this information separately does not pose a threat to the client's privacy, the combined information allows surprisingly high query recovery rates.

We consider different privacy-preserving SSE schemes that hide access pattern information and show how to adapt our attack against them.
Our evaluation confirms that two of these defenses fail at providing a significant level of protection even when they are configured for high privacy regimes.
The third defense is effective at hiding the query volume information, but even a small amount of frequency data (250 possibly repeated queries from the client, when there are $1\,000$ possible keywords) can provide non-trivial query recovery rates ($23\%$).

We hope that our work inspires researchers to find solutions that not only hide the access pattern leakage but also reduce the search pattern leakage, which we believe is paramount towards achieving effective privacy-preserving SSE schemes.

\section*{Acknowledgments}

We gratefully acknowledge the support of NSERC for grants RGPIN-05849, CRDPJ-531191, IRC-537591 and the Royal Bank of Canada for funding this research. 
This work benefited from the use of the CrySP RIPPLE Facility at the University of Waterloo.

\section*{Availability}

Our code is available at \url{https://github.com/simon-oya/USENIX21-sap-code}.

\bibliography{mybib}{}
\bibliographystyle{plain}

\appendix
\section{Results as Percentage of Distinct Keywords Recovered}

In Section~\ref{sec:eval}, we measure the attack accuracy as the percentage of queries correctly recovered.
In this section, for completeness, we report the accuracy of our experiments as the percentage of \emph{unique keywords} the attack correctly identifies.
We call this the \emph{unweighted accuracy}, since it is not weighted by the number of times the client queries for each keyword.

\begin{figure}[t]
	\centering
	\includegraphics[width=0.95\columnwidth]{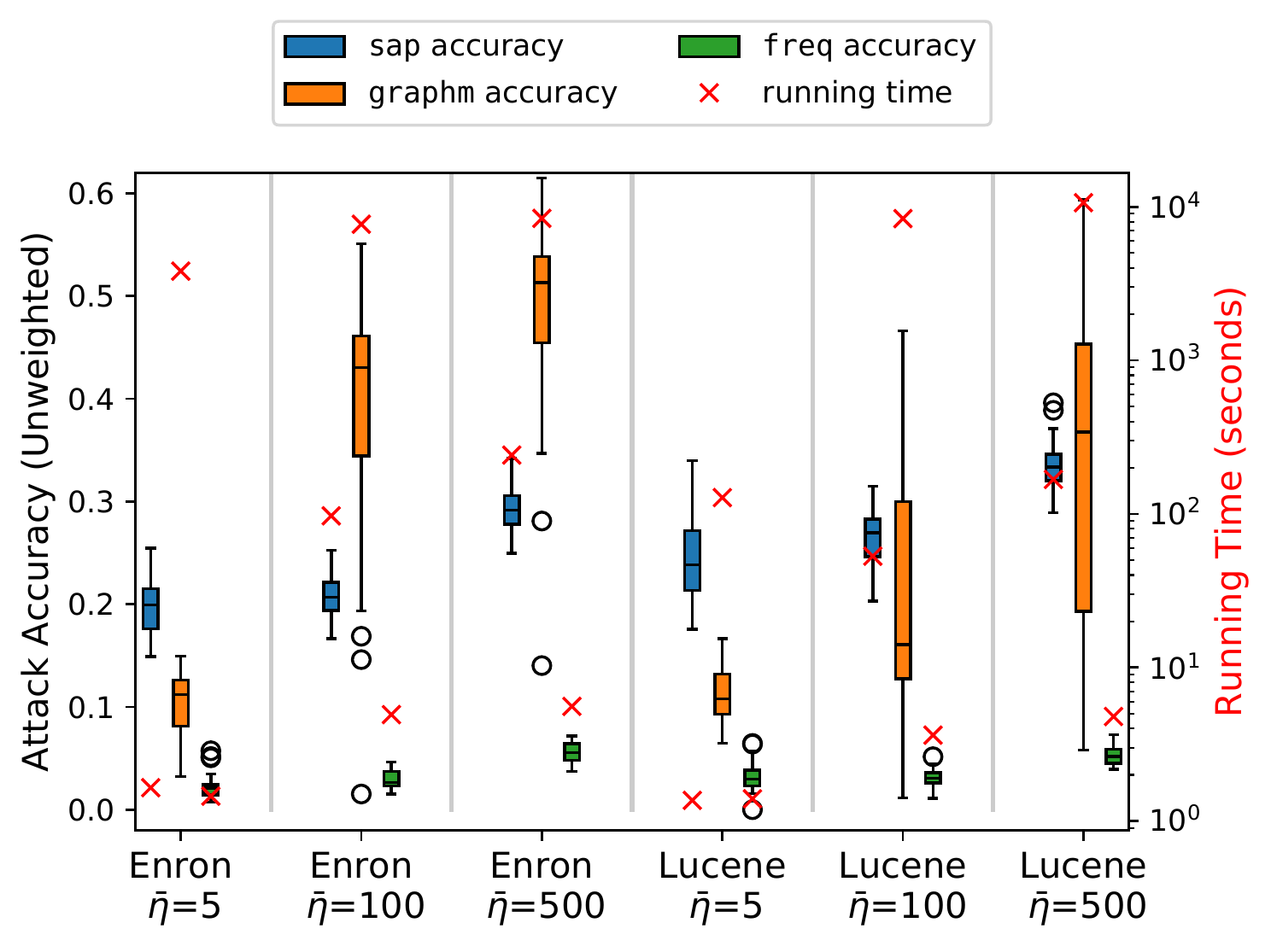}
	\caption{Unweighted recovery accuracy (boxes) and running time (\textcolor{red}{$\times$}) of attacks in different datasets with $\nqravg$ queries per week (50 weeks), with $\nkw=500$ keywords.}
\label{fig:attcomp_box_unique}
\end{figure}

Figure~\ref{fig:attcomp_box_unique} shows the comparison between attacks in terms of unweighted accuracy (regular accuracy in Figure~\ref{fig:attcomp_box} --- note the y-axes are different).
Both $\attack$ and $\liu$ achieve lower unweighted accuracy than regular (weighted) accuracy, since they are more likely to correctly recover queries corresponding to frequently queried keywords.
The unweighted accuracy of $\graphm$ is only slightly smaller than its regular accuracy; we conjecture this is because those keywords that are more popular in the dataset, and thus are easier to recover with co-occurrence information, are queried more often than unpopular keywords.
Even though $\graphm$ performs on average better than $\attack$ when the adversary observes a large number of queries, we note that $\graphm$ is still 1) computationally unfeasible for large keyword universe sizes, 2) performs worse than $\attack$ both in weighted and unweighted accuracies when the client performs few queries per week, and 3) completely fails against defenses such as PPYY~\cite{patel2019mitigating} and SEAL~\cite{demertzis2020seal}.

\begin{figure*}[t]
\begin{minipage}{\linewidth}
	\begin{minipage}{0.49\linewidth}
		\centering
		\includegraphics[width=0.95\linewidth]{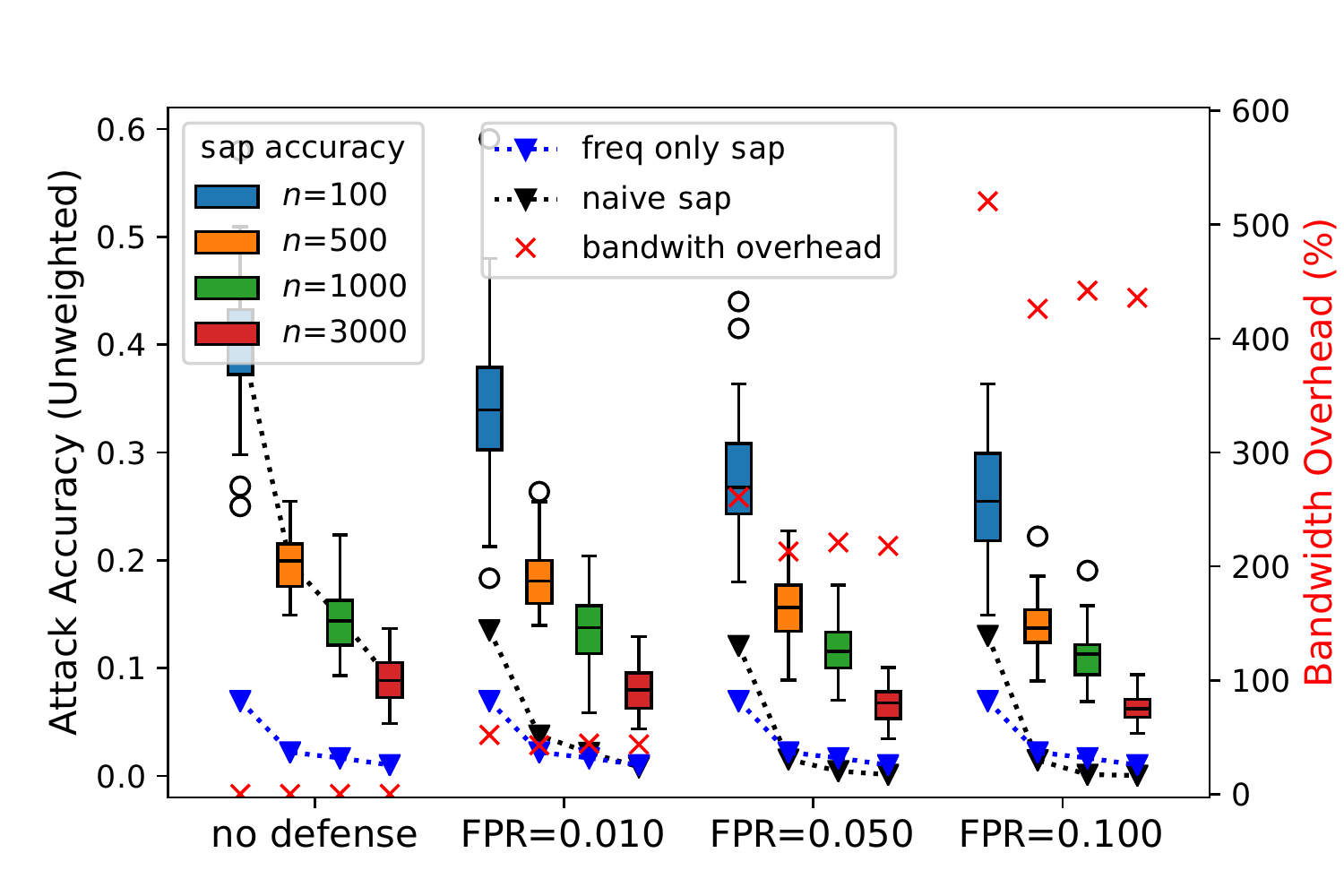}\\
		(a) Enron dataset
	\end{minipage} \hfill
	\begin{minipage}{0.49\linewidth}
		\centering
		\includegraphics[width=0.95\linewidth]{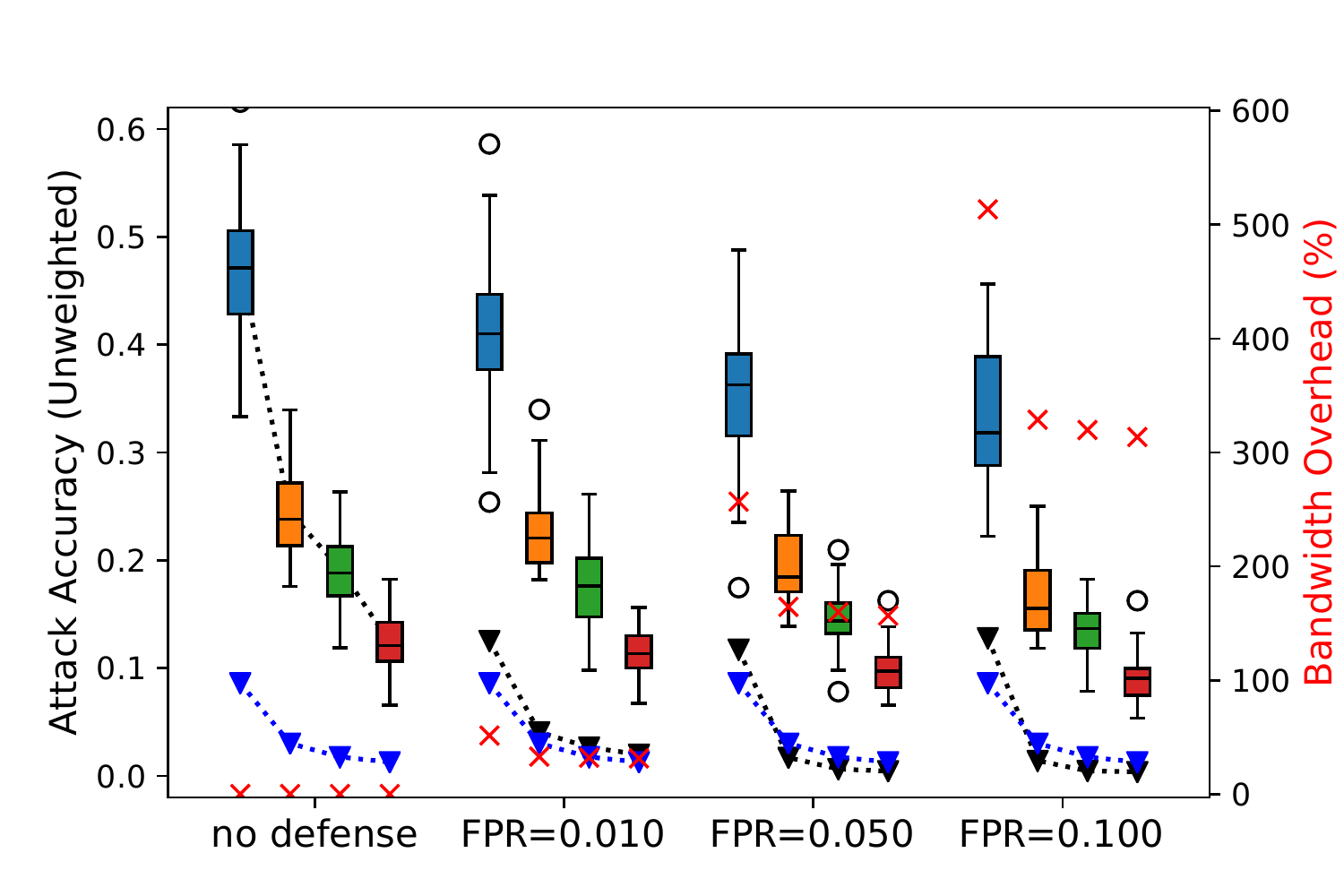}\\
		(b) Lucene dataset
	\end{minipage}
\end{minipage}
\caption{Unweighted accuracy of $\attack$ against CLRZ defense configured with $\TPR=0.999$ and varying $\FPR$ (50 weeks, $\nqravg=5$ queries/week).}
\label{fig:vs_clrz_unique}
\end{figure*}

\begin{figure*}[t]
\begin{minipage}{\linewidth}
	\begin{minipage}{0.49\linewidth}
		\centering
		\includegraphics[width=0.95\linewidth]{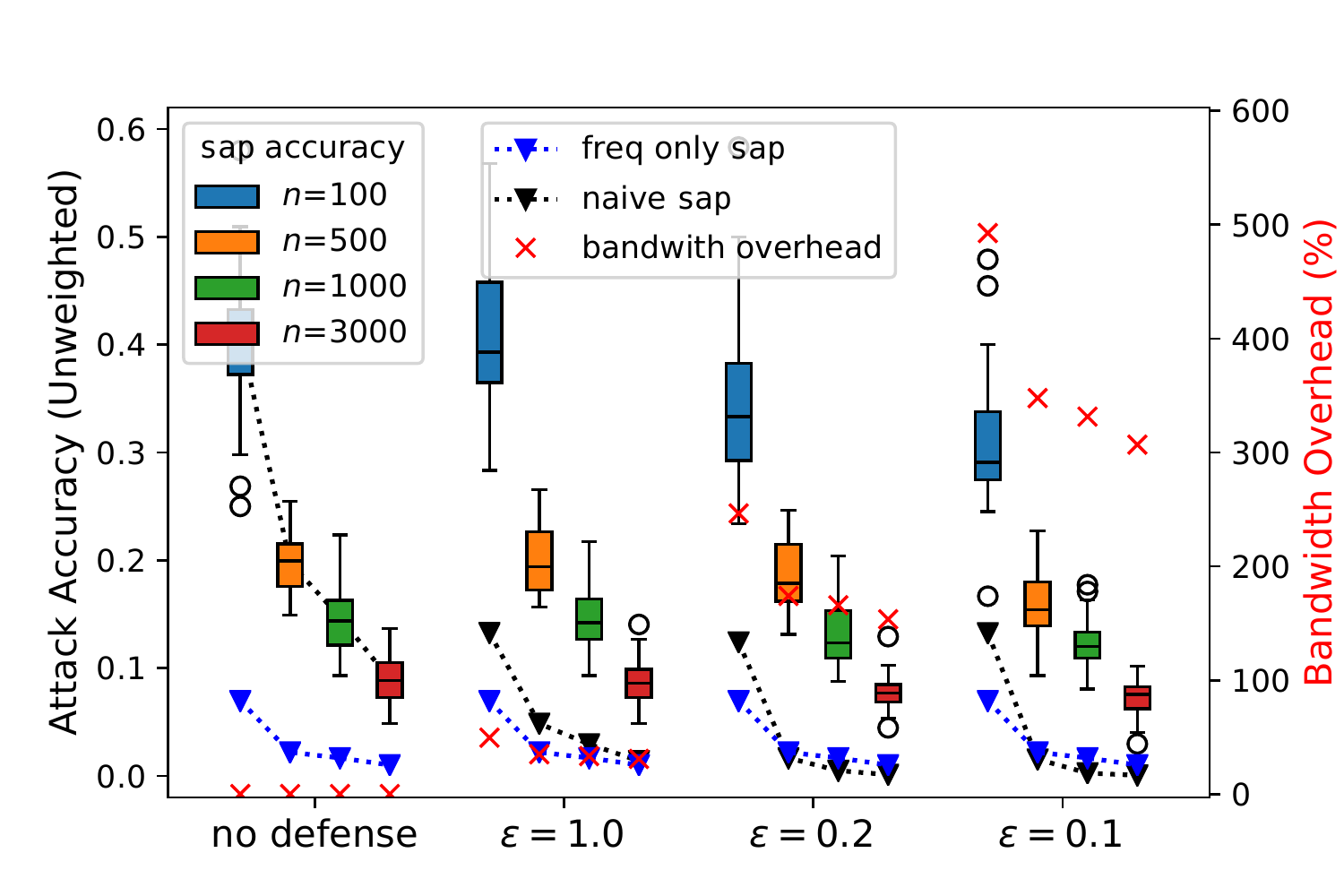}\\
		(a) Enron dataset
	\end{minipage} \hfill
	\begin{minipage}{0.49\linewidth}
		\centering
		\includegraphics[width=0.95\linewidth]{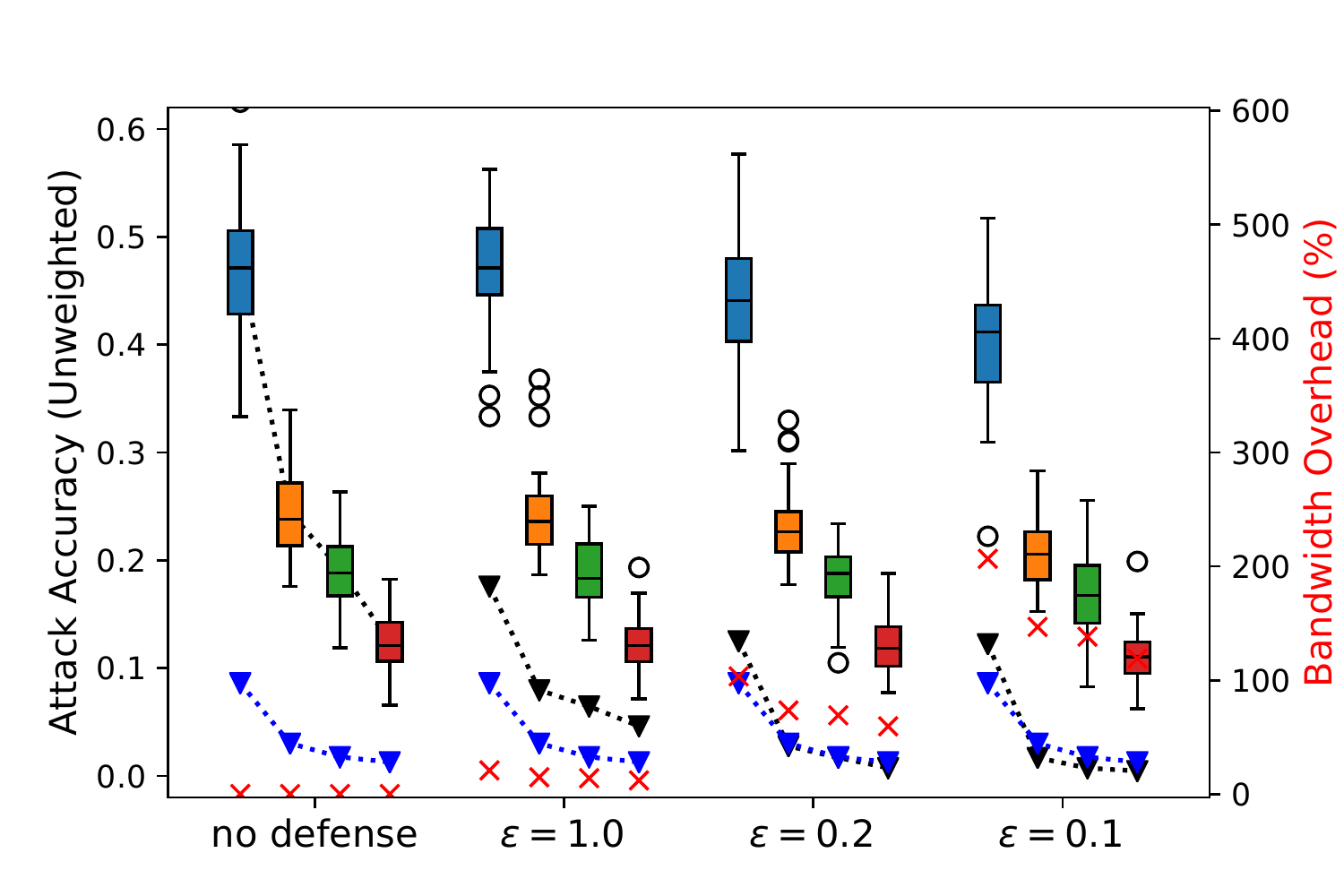}\\
		(b) Lucene dataset
	\end{minipage}
\end{minipage}
\caption{Unweighted accuracy of $\attack$ against PPYY defense with different privacy values $\epsilon$ (50 weeks, $\nqravg=5$ queries/week).}
\label{fig:vs_ppyy-unique}
\end{figure*}

\begin{figure*}[t]
\begin{minipage}{\linewidth}
	\begin{minipage}{0.49\linewidth}
		\centering
		\includegraphics[width=0.95\linewidth]{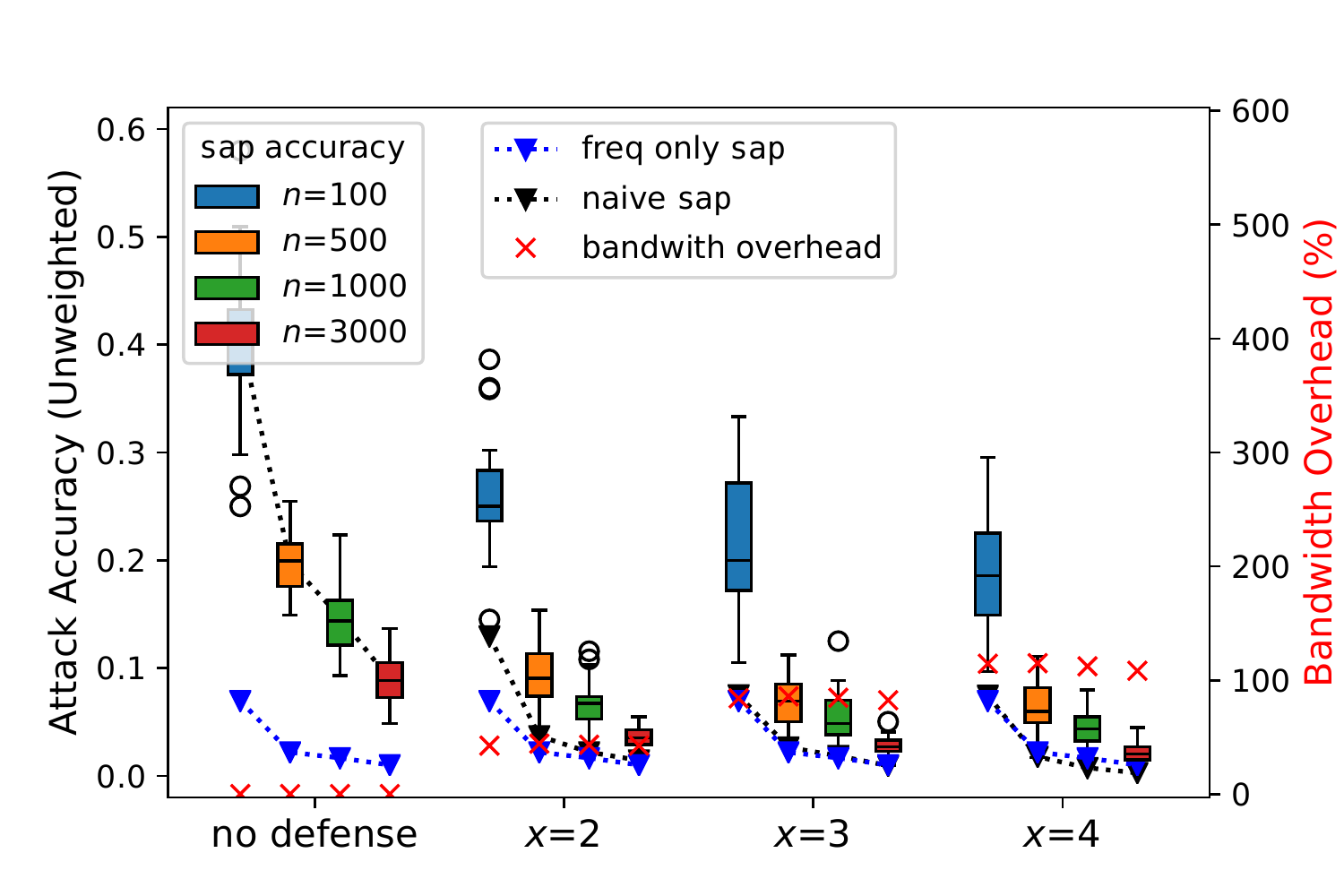}\\
		(a) Enron dataset
	\end{minipage} \hfill
	\begin{minipage}{0.49\linewidth}
		\centering
		\includegraphics[width=0.95\linewidth]{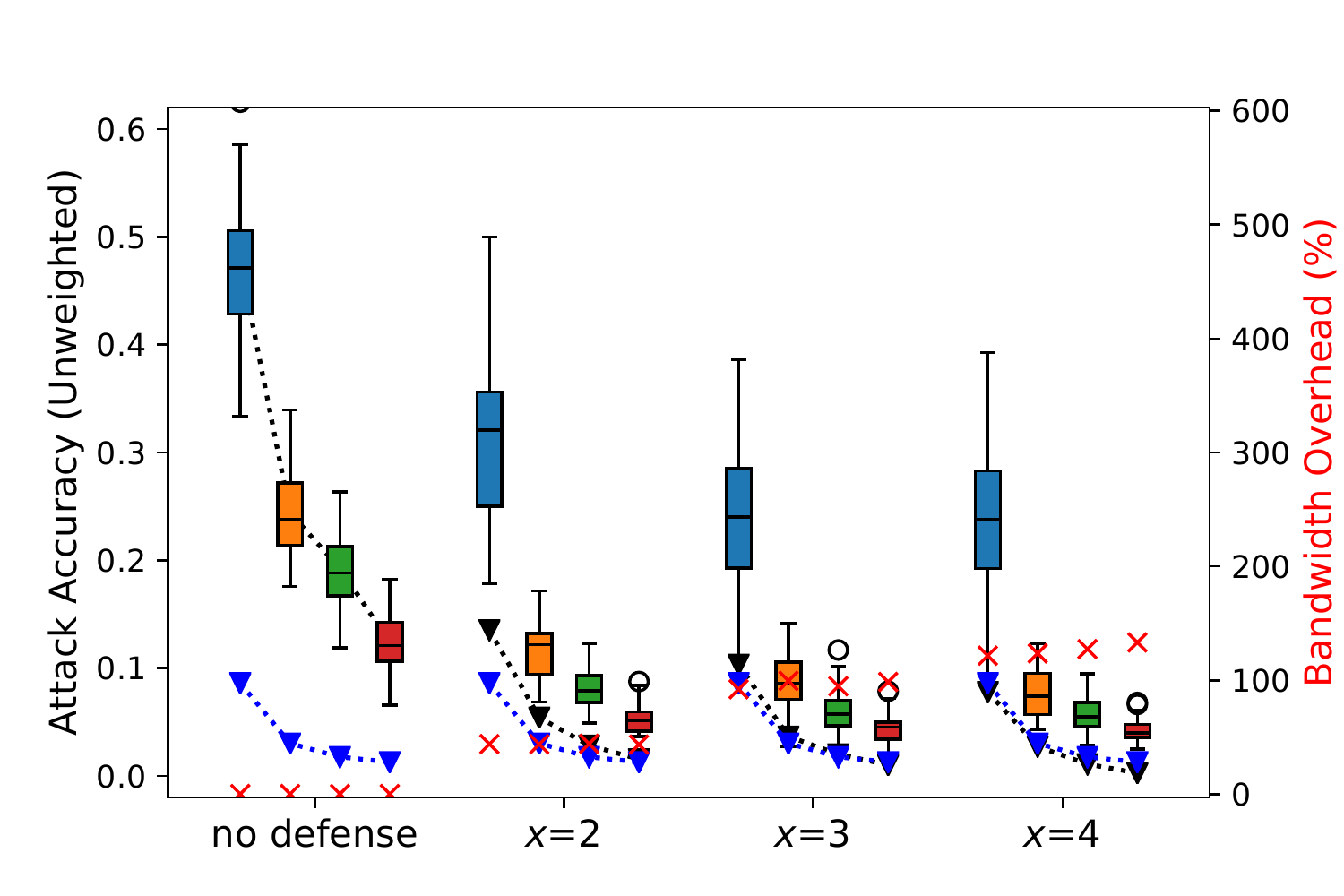}\\
		(b) Lucene dataset
	\end{minipage}
\end{minipage}
\caption{Unweighted accuracy of $\attack$ against SEAL defense for different values of multiplicative volume padding $x$ (50 weeks, $\nqravg=5$ queries/week).}
\label{fig:vs_seal_unique}
\end{figure*}

Figures~\ref{fig:vs_clrz_unique} to \ref{fig:vs_seal_unique} show the performance of $\attack$ in terms of the unweighted accuracy versus the three defenses we consider in the paper (the results for the regular accuracy are in Figures~\ref{fig:vs_clrz} to \ref{fig:vs_seal}).
Although the average number of unique keywords recovered by the attack is smaller than the average number of queries recovered, the results are qualitatively the same.

\end{document}